\documentclass[10pt,journal,compsoc]{IEEEtran}
\usepackage{hyperref,xcolor}
\usepackage{jabbrv}
\usepackage[utf8]{inputenc}
\usepackage[T1]{fontenc}
\usepackage{bm}
\usepackage[linesnumbered,ruled,vlined]{algorithm2e}
\usepackage[justification=justified, format=plain]{caption}
\usepackage{pgfplots}
\usepackage{multirow,multicol}
\usepackage{tabularx}
\usepackage{textgreek}
\usepackage{xfrac}
\usepackage{amstext}
\usepackage{amsmath}
\usepackage{amssymb}
\usepackage{dblfloatfix}
\usepackage{hhline}
\pgfplotsset{
    box plot/.style={
        /pgfplots/.cd,
        black,
        only marks,
        mark=-,
        mark size=\pgfkeysvalueof{/pgfplots/box plot width},
        /pgfplots/error bars/y dir=plus,
        /pgfplots/error bars/y explicit,
        /pgfplots/table/x index=\pgfkeysvalueof{/pgfplots/box plot x index},
    },
    box plot box/.style={
        /pgfplots/error bars/draw error bar/.code 2 args={%
            \draw  ##1 -- ++(\pgfkeysvalueof{/pgfplots/box plot width},0pt) |- ##2 -- ++(-\pgfkeysvalueof{/pgfplots/box plot width},0pt) |- ##1 -- cycle;
        },
        /pgfplots/table/.cd,
        y index=\pgfkeysvalueof{/pgfplots/box plot box top index},
        y error expr={
            \thisrowno{\pgfkeysvalueof{/pgfplots/box plot box bottom index}}
            - \thisrowno{\pgfkeysvalueof{/pgfplots/box plot box top index}}
        },
        /pgfplots/box plot
    },
    box plot top whisker/.style={
        /pgfplots/error bars/draw error bar/.code 2 args={%
            \pgfkeysgetvalue{/pgfplots/error bars/error mark}%
            {\pgfplotserrorbarsmark}%
            \pgfkeysgetvalue{/pgfplots/error bars/error mark options}%
            {\pgfplotserrorbarsmarkopts}%
            \path ##1 -- ##2;
        },
        /pgfplots/table/.cd,
        y index=\pgfkeysvalueof{/pgfplots/box plot whisker top index},
        y error expr={
            \thisrowno{\pgfkeysvalueof{/pgfplots/box plot box top index}}
            - \thisrowno{\pgfkeysvalueof{/pgfplots/box plot whisker top index}}
        },
        /pgfplots/box plot
    },
    box plot bottom whisker/.style={
        /pgfplots/error bars/draw error bar/.code 2 args={%
            \pgfkeysgetvalue{/pgfplots/error bars/error mark}%
            {\pgfplotserrorbarsmark}%
            \pgfkeysgetvalue{/pgfplots/error bars/error mark options}%
            {\pgfplotserrorbarsmarkopts}%
            \path ##1 -- ##2;
        },
        /pgfplots/table/.cd,
        y index=\pgfkeysvalueof{/pgfplots/box plot whisker bottom index},
        y error expr={
            \thisrowno{\pgfkeysvalueof{/pgfplots/box plot box bottom index}}
            - \thisrowno{\pgfkeysvalueof{/pgfplots/box plot whisker bottom index}}
        },
        /pgfplots/box plot
    },
    box plot median/.style={
        /pgfplots/box plot,
        /pgfplots/table/y index=\pgfkeysvalueof{/pgfplots/box plot median index}
    },
    box plot width/.initial=1em,
    box plot x index/.initial=0,
    box plot median index/.initial=3,
    box plot box top index/.initial=4,
    box plot box bottom index/.initial=2,
    box plot whisker top index/.initial=5,
    box plot whisker bottom index/.initial=1,
}

\newcommand{\boxplot}[2][]{
    \addplot [box plot median,#1] #2;
    \addplot [forget plot, box plot box,#1] #2;
    \addplot [forget plot, box plot top whisker,#1] #2;
    \addplot [forget plot, box plot bottom whisker,#1] #2;
}

\newcommand{\boxplotfilled}[2][]{
    \addplot [box plot median,#1, white, thick] #2;
    \addplot [forget plot, box plot box, fill,#1] #2;
    \addplot [forget plot, box plot top whisker,#1] #2;
    \addplot [forget plot, box plot bottom whisker,#1] #2;
}

\newcommand{\boxplotlegend}[2][]{
    \addlegendimage{#1}
    \addlegendentry{#2}
}
\pgfplotsset{compat=1.14}

\usetikzlibrary{patterns} 

\DeclareMathOperator*{\argmax}{arg\,max}

\ifCLASSOPTIONcompsoc
  \usepackage[nocompress]{cite}
\else
  \usepackage{cite}
\fi
\ifCLASSINFOpdf
\else
\fi
\begin{document}
%
\title{Reconstruction of Gene Regulatory Networks using Multiple Datasets}
%
%
%
%

\author{Mehrzad~Saremi,~M.Sc., Department of Artificial Intelligence at Amirkabir University of Technology
        Maryam~Amirmazlaghani, Assistant Professor, Department of Artificial Intelligence at Amirkabir University of Technology
\IEEEcompsocitemizethanks{\IEEEcompsocthanksitem M. Saremi and M. Amirmazlaghani are with Amirkabir University of Technology, Tehran, 1591634311, Iran.\protect\\
E-mail addresses: mehrzad.saremi@aut.ac.ir and mazlaghani@aut.ac.ir
}
}

\IEEEtitleabstractindextext{%
\begin{abstract}
\textbf{Motivation:} Laboratory gene regulatory data for a species are sporadic. Despite the abundance of gene regulatory network algorithms that employ single data sets, few algorithms can combine the vast but disperse sources of data and extract the potential information. With a motivation to compensate for this shortage, we developed an algorithm called GENEREF that can accumulate information from multiple types of data sets in an iterative manner, with each iteration boosting the performance of the prediction results. \\
\textbf{Results:} The algorithm is examined extensively on data extracted from the quintuple DREAM4 networks and DREAM5's Escherichia coli and Saccharomyces cerevisiae networks and sub-networks. Many single-dataset and multi-dataset algorithms were compared to test the performance of the algorithm. Results show that GENEREF surpasses non-ensemble state-of-the-art multi-perturbation algorithms on the selected networks and is competitive to present multiple-dataset algorithms. Specifically, it outperforms dynGENIE3 and is on par with iRafNet. Also, we argued that a scoring method solely based on the AUPR criterion would be more trustworthy than the traditional score.\\
\textbf{Availability:} The Python implementation along with the data sets and results can be downloaded from \url{github.com/msaremi/GENEREF}
\end{abstract}

}

\maketitle

\IEEEdisplaynontitleabstractindextext

%
\IEEEpeerreviewmaketitle

\IEEEraisesectionheading{\section{Introduction}\label{sec:introduction}}

%
%
%
%
\IEEEPARstart{G}{ene} Regulatory Networks (GRNs) model the crucial interactional patterns that control the molecular machinery in the cells of an organism \cite{barabasi:2004gene-regulatory-network}. With the advent of high-troughput technologies, simultaneous expression data from several genes have been available for many species and many algorithms have been proposed that can reconstruct GRNs from these data. Though algorithms perform quite well on artificial data sets, the shortcomings the algorithms encounter when dealing with laboratory data, keeps deciphering GRN data an open challenge in the field of systems biology. Multiplicity and variety of genomic data sets gathered for a species provide an opportunity to develop algorithms that employ multiple data sets to extract more information and yield more accurate results in the real world challenges. In this paper, we provide an algorithm, named GENEREF (GEne NEtwork inference with REgularized Forests), that brings together the idea of boosting in machine learning \cite{freund:1999boosting} and the capability of feature scoring in decision trees into a single model. The model is capable of using multiple types of data sets for the task of GRN reconstruction in arbitrary orders.

There are typically two types of transcriptome data that are used to reconstruct GRNs: Steady-state data and time series data. The steady-state data are obtained by applying simultaneous perturbations on a number of genes and measuring the expressions after the network reaches a steady state\cite{irrthum:2010genie3,guo:2016gene}. Data of this type is plentiful \cite{xiong:2012gene}. In the case of multiple perturbations, the system can be corresponding to profiles acquired from different donors or biological replicates \cite{irrthum:2010genie3}. The time series data on the other hand can in principle be more informative, since they capture the dynamic behaviour of the system throughout the time. Despite the temporal information, they are less available due to the capturing and timing difficulties in the process. The systematical issues that arise with regards to this type of data sets include rarity of fairly uniform cell populations to sample over time, de-synchronization of cells during the experiment, and high costs of dense sampling of the population \cite{bar:2012timeseries-experiments}.

In the realm of steady state data, ensemble methods based on regressor trees have shown to be one of the most effective approaches to extract the relation between the genes in a GRN \cite{irrthum:2010genie3,huynh:2015jump3,petralia:2015irafnet,geurts:2018dyngenie3}, with GENIE3 being the best performer in the DREAM4 challenge \cite{stolovitzky:2007dream4challenge}. The DREAM4 challenge measures performance of variant algorithms on the reverse engineering of five steady state \textit{in silico} networks. GENIE3 divides a problem with $g$ genes into $g$ sub-problems and proceeds to solve each problem by the intrinsic feature ranking property of random forests or extra trees. This work can be thought of as a multi-iteration algorithm where each iteration contains an extension to the GENIE3 algorithm applied on a data set. Unlike GENIE3, it can exploit multiple data sets and is less prone to over-fitting because of the regularization mechanisms it employs.

An advent algorithm that we will consider as the competing algorithm with our work is dynGENIE3. It is another extension to GENIE3, that introduces the differential equations governing the gene expression relationships to the typical tree-based ensemble methods used in GENIE3. The differential equations are formed based on the assumption that the transcription rate of the target gene is a function of the expression levels of all other genes and the decay rate of the products of the target gene itself. Compared to dynGENIE3, two advantages of our algorithm are that it does not depend on the decay rate parameters included in the differential equations and it can be extended to use different types of steady state data (e.g. double knock-out and multi-perturbation) \cite{geurts:2018dyngenie3}.

In sum, GENEREF is a highly scalable method that consistently increases the performance of the base algorithm (GENIE3), and outperforms the present multi-perturbation GRN algorithms. The algorithm also receives better scores than the competing algorithms (dynGENIE3 and iRafNet) on some networks when using the typical scoring method incorporating both AUROC and AUPR. Moreover, GENEREF drastically outperforms dynGENIE3 when a new scoring method based only on the AUPR metric is used.

In what follows we will introduce our method. This paper also encompasses the evaluation of our algorithm on the \textrm{in silico} DREAM4 networks in comparison to other network reverse-engineering methods.

\section{Methods}
The problem of GRN reconstruction can be summarized as initially inferring rankings for the potential regulatory links from the data and eventually applying a threshold on these rankings to obtain a predicted network. GENEREF -- along with many other algorithms in the field \cite{irrthum:2010genie3,haury:2012tigress,geurts:2018dyngenie3,zheng:2018bixgboost} -- deals only with the former issue and puts the latter aside. Although, it is possible to compute the expected optimal threshold cut-off \cite{grimaldi:2011regnann}, comparable algorithms typically satisfy with taking into account all of the distinguishing thresholds when reporting their evaluation results. (We will discuss these evaluations metrics further in sub-section \ref{subsection:evaluation}.) In what follows in this section, we provide the details of problem and the preliminary structures of the algorithm, along with its definition.

\subsection{Network Inference with Tree-based Methods}
Tree-based algorithms belong to the group of GRN inference methods that associate a value to each ordered pair of genes predicting how certainly there is a link in the ground truth network from the first to the second gene. This is done by finding a regression that predicts the expression of one gene based on the others. The basic idea is to decompose the problem of finding the regulatory links in $g$ genes into $g$ sub-problems. Each sub-problem will then be a regression problem in which the goal is to find the best prediction of the target gene based on the expression values of all the other genes.

Generally, tree-based methods -- be them random forest or extra trees -- are non-parametric algorithms that implement the idea of regression algorithms using regression trees: For each gene $G_j$ ($j = 1, ... , g$), a sub-problem is defined. In the $j$-th sub-problem, the expression values of the $j$-th gene will be considered as the target values and the expression values of all other genes will express the feature values. A regression is then solved for each sub-problem using a tree-based method.

Trees have the innate capability of calculating feature importance scores. As the importance of the $i$-th feature in the $j$-th sub-problem gives an estimation about the certainty of the existence of an edge from gene $G_i$ to $G_j$, the tree-based algorithms merge these scores measured for each sub-problem to produce a final ranking for all edges in the network.

In a random forest setup, the average of the importance values of all features for a tree is close to the variance of the output variable. Therefore, to make the feature importances in all sub-problems homogeneous, the expression levels of all genes are typically normalized before the data is provided to the algorithm \cite{irrthum:2010genie3}. In our work, we applied this pre-processing step for each data set before it is fed to a random forest.

\subsection{Regularized Random Forests}
Regularized Random Forests (RRFs) are an extension to the typical random forests that are superior to the regular forests in handling the issue of feature redundancy. When used as a feature selection technique, RRFs can run with the same setup as regular forests. Firstly introduced in \citen{deng:2012rrf}, RRFs addressed the feature redundancy issue in the problem of feature selection by setting a penalty against adding new features to the currently selected features set and trying to maintain a minimal feature set.

In the original RRF framework, consider a feature selection problem with $g$ features represented as the feature set $F = \left\{z_j\mid j = 1,...,g\right\}$. Trees in a forest are formed in a top-down manner, selecting one feature and a split point for each node that is being added to the hierarchy of the tree. The selected feature and its split point will determine the importance of each feature after the construction of all trees. As a tree in the forest is built, there will be more and more features based on which the learning data is split. We name the set of these features $F_\text{S}$. In every node of each tree, the node is configured based on the improvement function as follows:
\begin{equation}
\text{Improvement}_\text{R}(z_j, \aleph) := 
    \begin{cases}
        \lambda\text{Improvement}(z_j, \aleph) & z_j\not\in F_\text{S}\\
        \text{Improvement}(z_j, \aleph) & \text{otherwise.}
    \end{cases}
\label{eq:improvement}
\end{equation}
where $\aleph$ is the node being configured and $\lambda$ is a regularization parameter in the range $\left[0, 1\right)$. $\text{Improvement}(.)$ is the function used in regular random forests and in the case of a regression problem, can be any of the Mean Squared Error ($\mathit{MSE}$) or Mean Absolute Error ($\mathit{MAE}$) gains \cite{breiman:2017cart}. The $\text{Improvement}_\text{R}$ function is tested for a randomly selected set of features and their randomly selected split points and the feature with the maximum improvement will be assigned to the node:
\begin{equation}
z_\aleph = \argmax_{z_j}\left\{ \text{Improvement}_\text{R}(z_j, \aleph)\right\}
\end{equation}
where $z_\aleph$ is the feature assigned to node $\aleph$. $F_\text{S}$ is then updated by the insertion of $z_\aleph$. It can be seen that as $\lambda$ shrinks in equation \ref{eq:improvement}, the tree biases more towards selecting features that are already in the selected feature set, and hence will be more picky about choosing features in computed nodes. By choosing the proper value for $\lambda$, it will therefore become possible to omit the redundant features from the final selected features using an importance score.

After the construction of the whole forest, the relative importance of a feature $z_j$ is measured by averaging the values of the $\text{Improvement}$ function over all the nodes split by that feature:
\begin{equation}
\text{Importance}(z_j) = \frac{1}{\left|\mathbb{N}_{z_j}\right|} \sum_{\aleph \in \mathbb{N}_{z_j}} \text{Improvement}_\text{R}(z_j, \aleph)
\label{eq:importance}
\end{equation}
where $\mathbb{N}_{z_j}$ is the set of all nodes in the whole forest with the $z_j$ feature. These importance measurements guide the procedure of feature selection in the RRF.

Empirical data indicate that GRNs are sparse networks with in-degree distribution concentrated around the mean connectivity \cite{thieffry:1998ecoli-indegree, shen:2002ecoli-indegree2, guelzim:2002yeast-indegree}. We speculate that the prediction performance of such networks can be enhanced using regularization methods. The regularization also reduces the probability of over-fitting. Based on these grounds, we utilize the same idea of RRFs and use an algorithm similar to the guided regularized random forests proposed in \citen{deng:2013grrf}. We propose an iterative algorithm that uses various sets of learning data to reconstruct the final regulatory network. Each of these data sets can be seen as the output of an experiment. This work also inherits the decomposition approach used in many GRN reverse engineering algorithms like GENIE3 \cite{irrthum:2010genie3} and TIGRESS \cite{haury:2012tigress}. On each iteration, the algorithm picks a data set and uses the regularization parameters obtained from the previous iteration to guide a random forest. At the end, the network edges are ranked based on the final feature importance values.

\subsection{Problem Definition}
Suppose that we want to discover the regulatory links between $g$ genes ($G_1, ..., G_g$) of a certain species. Also, consider there to be $M$ learning data sets $\mathbf{X}_l$, ($l = 1, ..., M$) of that species. Each data set represents a set of gene expression profiles obtained from a corresponding experiment. The $l$-th data set is comprised of $N_l$ gene expression profiles. Each gene profile is a snapshot containing the expression values of all of the $g$ genes.

Here, we consider two types of experiments: steady state experiments and time-series experiments. To differentiate between the type of data sets that are generated using these experiments, we superscribe them with appropriate abbreviations. In our case, every arbitrary data set $\mathbf{X}_l$ is either represented as $\mathbf{X}_l^{\left(\text{SS}\right)}$ if it holds the measurements of a steady state experiment, or as $\mathbf{X}_l^{\left(\text{TS}\right)}$ if it holds the measurements of a time-series experiment.

In a steady state experiment, perturbations are applied to randomly selected subset of genes (i.e. their basal expression is modified) and snapshots are captured after the network reaches a steady state. Each element in a steady state data set corresponds to the expression values of all genes in one of these snapshots. If the experiment producing $\mathbf{X}_l^{\left(\text{SS}\right)}$ entails $N_l$ profiles labeled with $e_{1}, ..., e_{N_l}$, then $\mathbf{X}_l^{\left(\text{SS}\right)}$ will be comprised of the following elements:
\begin{equation}
\mathbf{X}_l^{\left(\text{SS}\right)} = \left\{ \mathbf{x}_l(e_{1}), \mathbf{x}_l(e_{2}), ..., \mathbf{x}_l(e_{N_l}) \right\}
\text{\,,}
\end{equation}
where $\mathbf{x}_l(e_{i})$ is a vector containing the expression levels of all $g$ genes in profile $e_{i}$ in that experiment:
\begin{equation}
\mathbf{x}_l(e_{i}) = \left[ x_{l,1}(e_{i}), x_{l,2}(e_{i}), ..., x_{l,g}(e_{i}) \right]^T
\text{\,,}
\end{equation}
In this vector, $x_{l,j}(e_{i})$ is the expression level of gene $G_j$ in profile $e_{i}$. We also denote the vector of the expression level of all genes except $G_j$ in profile $e_{i}$ using $\mathbf{x}_{l,-j}(e_{i})$:
\begin{equation}
\mathbf{x}_{l,-j}(e_{i}) = \left[ ...,  x_{l,j-1}(e_{i}), x_{l,j+1}(e_{i}), ... \right]^T
\end{equation}

In a time series experiment, interventions are applied to a subset of genes, and while their causal effect is propagated through the GRN, the expression of all genes are probed simultaneously at specific time points. Consider that the profiles in the experiment producing $\mathbf{X}_l^{\left(\text{TS}\right)}$ are captured at time points $t_{l,1}, ..., t_{l,N_l}$ after the perturbation is applied ($i < j$ indicates $t_{l,i} < t_{l,j}$). $\mathbf{X}_l^{\left(\text{TS}\right)}$ will consist of $N_l$ elements ordered as follows:
\begin{equation}
\mathbf{X}_{l}^{\left(\text{TS}\right)} = \left\{ \mathbf{x}_l(t_{l,1}), \mathbf{x}_l(t_{l,2}), ..., \mathbf{x}_l(t_{l,N_l}) \right\}
\text{\,,}
\end{equation}
where $\mathbf{x}_l(t_{l,i})$ is a vector that contains the expression levels of all $g$ genes at time $t_{l,i}$:
\begin{equation}
\mathbf{x}_l(t_{l,i}) = \left[ x_{l,1}(t_{l,i}), x_{l,2}(t_{l,i}), ..., x_{l,g}(t_{l,i}) \right]^T
\end{equation}

For each multi-factorial perturbation data set, a learning problem composed of $g$ sub-problems can be constructed, such that in the $j$-th sub-problem, the $j$-th gene values become the target and all the other values constitute the input features. Every sub-problem is parallel to a regression problem where the goal is to find functions $f_{l,j}(.)$ that minimizes the errors $\epsilon_{l,j}$ in the following equation:
\begin{equation}
x_{l,j}(e_i) = f_{l,j}(\mathbf{x}_{l,-j}(e_i)) + \epsilon_{l,j}(i), \quad \forall i
\text{,}
\label{eq:ss_function}
\end{equation}
and
\begin{equation}
\epsilon_{l,j} = \sum_{i=1}^{N_l} \epsilon_{l,j}(i)
\end{equation}
Note that in the steady-state setup the value of $G_j$ is excluded from the domain of function $f_{l,j}$, because steady-state data sets have no information regarding causal self-loops.

Time-series learning data can be treated similarly, with the difference that the causal effect of the regulators appear after a time delay. Hence, the expression of each gene will be a function of the expression of all genes (possibly including the same gene) in a previous time point. This is equivalent to an auto-regressive model \cite{michailidis:2013autoregressive}. We use the same expression as equation \ref{eq:ss_function} with the exception that the input features will be chosen from a previous time point $t_{i-k}$ ($k > 0$):
\begin{equation}
x_{l,j}(t_i) = f_{l,j}(\mathbf{x}_{l}(t_{l,i-k})) + \epsilon_{l,j}(i), \quad \forall i
\end{equation}
For the sake of simplicity, we will keep $k = 1$ throughout this paper.

These regression sub-problems can be solved by the regular random forests or the guided RRFs within which the importance of a feature will be measured using equation \ref{eq:importance}. When the guided RRFs are used, the regularization parameters should be provided as introduced in sub-section \ref{the-generef-framework}. Throughout the rest of this paper we will drop the superscript from the notation of data sets wherever the type of the data set is not important.

\begin{figure*}[t!]
\centering
\includegraphics[width=0.9\linewidth]{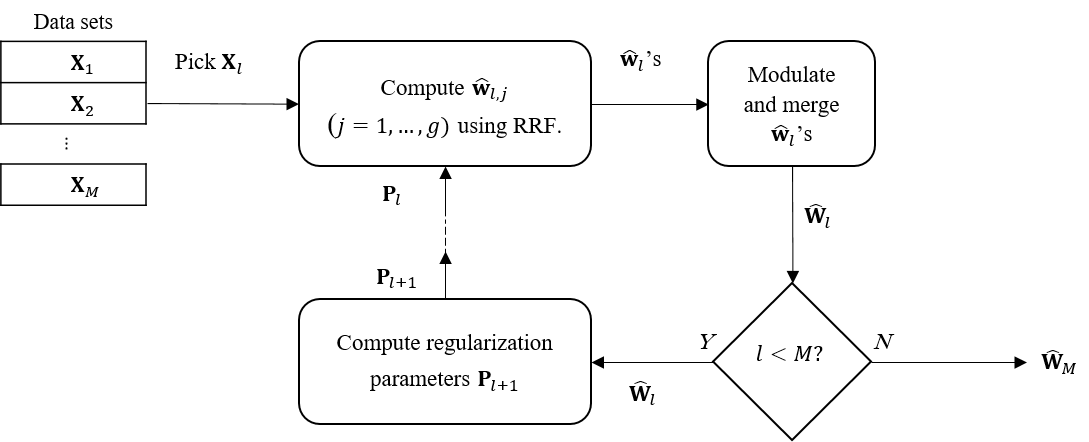}
\caption{Folded data flow diagram of GENEREF. On the $l$-th iteration, the $l$-th data set $\mathbf{X}_l$ and the $l$-th regularization matrix $\mathbf{P}_l$ are fed to a random forest and the $l$-th confidence matrix $\hat{\mathbf{W}}_l$ will be produced. The next level's regularization matrix $\mathbf{P}_{l+1}$ is then constructed based on $\hat{\mathbf{W}}_l$.}
\label{fig:generef}
\end{figure*}

\subsection{The GENEREF framework}
\label{the-generef-framework}
The goal of a multi-dataset algorithm is to combine available data sets in a plausible manner to produce a network as similar to the ground-truth network as possible. (We will discuss the similarity criteria in sub-section \ref{subsection:evaluation}.) Here, we represent the ground-truth network by an adjacency matrix $\mathbf{W}_{g\times g}$. Our algorithm, GENEREF, produces \textit{confidence matrices}, $\hat{\mathbf{W}}_l$ ($l \in \left\{1, ..., M\right\}$), that are of the same size as $\mathbf{W}$. The elements of the last confidence matrix, $\hat{\mathbf{W}}_M$, are expected to show how certainly the corresponding edges will be appearing on the actual network. The goal of GENEREF is therefore to produce a $\hat{\mathbf{W}}_M$ on which the elements corresponding to the real connections of the network get higher values than the others.

GENEREF works in an iterative manner: On each iteration it takes a pre-processed\footnotemark\ data set $\mathbf{X}_l$ and provides a set of regressors (here, the regular or regularized random forests) with it. Figure \ref{fig:generef} shows the overall structure of the algorithm. There are $M$ data sets $\mathbf{X}_1, ..., \mathbf{X}_M$ in total. The data sets can be of either type. There is also no restriction upon the uniqueness of the data sets. Two data sets, $\mathbf{X}_i$ and $\mathbf{X}_j$, are allowed to share elements or be equal.
In principle, the algorithm accepts any order of data sets.
\footnotetext{The pre-processing is done as in the GENIE3 algorithm \cite{irrthum:2010genie3}.}

On each iteration, GENEREF picks a data set and decomposes it to $g$ new data sets, each of which is represented as a separate sub-problem to be solved by a regular or regularized random forest. On the first iteration, a set of $g$ regular random forests are trained with the decompositions of data set $\mathbf{X}_{1}$. Consider sub-problem $j$ on this iteration. The random forest regressor solves this sub-problem and provides importance scores for the features. We represent these scores as a vector $\hat{\mathbf{w}}_{1,j} = \left[\hat{w}_{1,j,1}, ...,  \hat{w}_{1,j,j}, ... , \hat{w}_{1,j,g}\right]^T$ where $\hat{w}_{1,j,i}$ is the importance score for the $i$-th feature in the $j$-th sub-problem of data set $\mathbf{X}_1$ and is caculated based on equation \ref{eq:importance}. Note that in the steady state configuration, the $j$-th feature is not present and therefore $\hat{w}_{l,j,j} = 0$. However, in a time-series data set, the $j$-th feature can be of a higher value.

Up to this stage, our algorithm parallels GENIE3 \cite{irrthum:2010genie3}. GENIE3 generates a temporary $\hat{\mathbf{W}}$ matrix by simply merging the column vectors $\hat{\mathbf{w}}_{1,j}$, and represents it as the final edge confidence matrix. The issue with this approach is that since the sub-problems have been solved independently, feature importance values will not be comparable across the columns and cannot be joined without further modification. Therefore, a \textit{modulation} phase is needed on the columns before joining them. 
An ideal regressor is expected to assign the highest possible value (i.e. $1$) to the ``relevant'' features and the least possible value (i.e. $0$) to the ``irrelevant'' ones. By diversifying the values on each column to these extents, the modulation phase helps the potential links on various columns to be treated equally. While GENIE3 keeps $\hat{\mathbf{W}}$ as is and reports it as the final confidence matrix, GENEREF goes on to performing the modulation phase by applying the following standardization function on each column of $\hat{\mathbf{W}}$:
\begin{equation}
{\hat{\mathbf{w}}_{1,j}}' = \frac
{\hat{\mathbf{w}}_{1,j}' - \min{\left(\hat{\mathbf{w}}_{1,j}'\right)}}
{\max{\left(\hat{\mathbf{w}}_{1,j}'\right)} - \min{\left(\hat{\mathbf{w}}_{1,j}'\right)} + 
\text{\straightepsilon}}
\text{,}
\label{eq:modulation}
\end{equation}
where $\text{\straightepsilon}$ is a small constant. 

The edge confidence matrix $\hat{\mathbf{W}}_{1}$ is constructed by merging the column vectors ${\hat{\mathbf{w}}_{1,j}}'$, $j=1,...,g$:
\begin{equation}
\hat{\mathbf{W}}_{1} = \left[{\hat{\mathbf{w}}_{1,1}}', ..., {\hat{\mathbf{w}}_{1,g}}'\right]
\label{eq:construct_W}
\end{equation}

\begin{algorithm*}[t!]
\SetAlgoLined
\DontPrintSemicolon
\SetKwInOut{Input}{inputs}
\SetKwInOut{Output}{output}
\SetKwProg{Fn}{begin}{}{}
\Input{Number of data sets $M$\newline Set of all data sets $\left\{\mathbf{X}_l\middle|l = 1,...,M\right\}$}
\Output{The edge confidence values matrix $\hat{\mathbf{W}}_M$}
\Fn{}{
Set $l=1$\;
 \While{$l \leq M$}{
  Take $\mathbf{X}_l$ as the current data set and normalize it\;
  \eIf{$l=1$}{
    Compute ${\hat{\mathbf{w}}_{l,j}}$'s ($j=1, ..., g$) from $\mathbf{X}_l$ using the regular random forest\;
  }{
   Compute ${\hat{\mathbf{w}}_{l,j}}$'s ($j=1, ..., g$) from $\mathbf{X}_l$ and $\mathbf{P}_{l}$ using the regularized random forest\;
  }
  Apply the modulation phase on ${\hat{\mathbf{w}}_{l,j}}$'s and produce ${\hat{\mathbf{w}}_{l,j}}'$'s ($j=1, ..., g$) using equation \ref{eq:modulation}\;
  Construct $\hat{\mathbf{W}}_l$ by merging ${\hat{\mathbf{w}}_{l,j}}'$ ($j=1, ..., g$) using equation \ref{eq:construct_W}\;
  \If{$l=M$}{
    \Return $\hat{\mathbf{W}}_l$ as the final scorings\;
  }{}
  Compute the regularization parameters matrix $\mathbf{P}_{l+1}$ using equations \ref{eq:remove-relative-importances} and \ref{eq:regularization}\;
  $l = l + 1$
 }}
 \textbf{end}
 \caption{The GENEREF algorithm.}
 \label{algorithm:generef}
\end{algorithm*}

On each of the next iterations, GENEREF exploits the next data set. The feature importance values are computed using the regularized random forest. However, since on these iterations GENEREF has already acquired knowledge about which features are potentially more important using the previous learning data, it uses this knowledge to regularize the random forests. To this end, GENEREF uses a vector of regularization values for each sub-problem at each iteration. We represent these vectors by $\mathbf{p}_{l,i} = \left[\rho_{1,i,1}, ..., \rho_{1,i,g}\right]^T$ ($i \in \left\{1, ..., g\right\}$), where $\rho_{1,i,j}$ is the regularization value of the $j$-th feature in the $i$-th subproblem on iteration $l$. The $\mathbf{p}_{l,i}$ vectors form the $l$-th iteration's regularization matrix using the following equation:
\begin{equation}
\mathbf{P}_l = [\mathbf{p}_{l,1}, ..., \mathbf{p}_{l,g}]
\text{.}
\end{equation}
The details about how the regularization parameters are computed are skipped for now and will be provided in sub-section \ref{regularization-parameters}. After these parameters are calculated for the current iteration, the RRF computes the improvement function for each node in the forest using equation \ref{eq:improvement2} and retrieves the new feature importance scores $\hat{\mathbf{W}}_{2}$.
\begin{equation}
\text{Improvement}_\text{R}(z_{i,j}, \aleph) = 
        \rho_{l,i,j}\text{Improvement}(z_{i,j}, \aleph)
\text{.}
\label{eq:improvement2}
\end{equation}
Here, $z_{i,j}$ is equivalent to the feature corresponding to the $j$-th gene in the $i$-th sub-problem.

GENEREF provides a $\hat{\mathbf{W}}_{l}$ matrix for each iteration in the same way as the second iteration. Once the algorithm finishes iteration $M$, it returns the final edge confidence matrix $\hat{\mathbf{W}}_{M}$. Every element $\hat{w}_{M,i,j}'$ on this matrix, shows the certainty that an edge exists from gene $G_i$ to gene $G_j$. Please refer to algorithm \ref{algorithm:generef} for the concise definition of this method. Standard performance criteria, as described in section \ref{section:results}, can be used to evaluate this matrix by comparing it to the adjacency matrix $\mathbf{W}$.

\subsubsection{Regularization parameters}
\label{regularization-parameters}
The regularization parameters matrix used on the $l$-th iteration of GENEREF, $\mathbf{P}_l$, is calculated using the edge confidence matrix values obtained in iteration $(l-1)$ in the algorithm, $\hat{\mathbf{W}}_{l-1}$. In this sub-section, we will introduce a method to properly convert $\hat{\mathbf{W}}_{l-1}$ to $\mathbf{P}_l$. This method encompasses an operation followed by a mapping. We refer to the mapping as the \emph{regularization mapping} throughout the rest of this paper.

The more the certainty of the presence of an edge, the more potentially important that edge will be, and therefore it should be penalized less.  Although the confidence values provide an ordering for the importance of each edge, there is no guarantee that their values depict the relative importance of the edges. In fact, using the random forest regressors is not a proper means to get accurate regularization values. Instead, we remove the relative importance values while keeping their ordering. The following function does this.

\begin{equation}
\omega_{l,i,j} = \frac{1}{g^2-1} \sum_{m,n \in \left\{1,...,g\right\}} \mathbf{1}(\hat{w}_{l,m,n} < \hat{w}_{l,i,j})
\label{eq:remove-relative-importances}
\end{equation}
where $\mathbf{1}(\cdot)$ is the indicator function. The operation helps us make sure that the whole domain of the mapping takes its effect evenly at each iteration. This gives a full control of how we want to take into account different edges based on their importance.

It has been reported that GRN's have fast decreasing in-degree distributions \cite{guelzim:2002topological,nair:2015improving}. This means that for each node there are only a few direct regulatory genes. Also, it has been shown that GRN's typically have few \textit{global} regulatory genes \cite{guelzim:2002topological,roy:2015ecoli-characterization,lawson:2004yeast-catabolite,nair:2015improving}. It is therefore expected that a few links should be preserved as the important ones and all the others should be left as unimportant. The important ones are preferably those with the highest importance values. It is the regularization mapping that determines the regularization values based on the retrieved importance of the edges.

For the regularization mapping, we chose a simple, yet fairly flexible function -- the Kumaraswamy cumulative distribution function (CDF) \cite{kumaraswamy:1980generalized}.
This extension of the power function, ensures selecting the few global regulators if proper parameters are set. However, any other non-decreasing functions that try to satisfy the aforementioned properties of GRNs can be tested too. We denote the Kumaraswamy CDF function using $\text{KCDF}\left(\cdot\right)$:
\begin{equation}
\text{KCDF}\left(\omega; \alpha, \beta\right) := 
1 - \left(1 - \left(\omega\right)^\alpha\right)^\beta
\text{.}
\end{equation}
Consider that $\bm{\Omega}_l$ is the matrix comprised of the elements $\omega_{l, i, j}$. The Kumaraswamy CDF alters the importance values by using two constant shape parameters $\alpha$ and $\beta$ and returns the regularization parameters matrix:
\begin{equation}
\mathbf{P}_{l} = \text{KCDF}\left(\bm{\Omega}_l; \alpha, \beta\right)
\text{.}
\label{eq:regularization}
\end{equation}
Both $\alpha$ and $\beta$ range from $0$ to $1$. Roughly speaking, greater $\alpha$ values extend the domain of non-regulatory edges, while greater $\beta$ values favour preserving a broader domain for regulatory links. When ${\alpha}  \xrightarrow{} 0$, all features are treated equally, there will be no regularization, and the RRF treats all features as equally important. 
We describe how selective the regularization mapping is by defining a property called the ``\emph{selectivity}'' of the mapping. In section \ref{section:results}, we will make use of this property to discuss the optimal value of $\alpha$ and $\beta$.

We define the selectivity property as the area between the regularization mapping and the line $y = 1$ and denote it by $\text{Sel}\left(\cdot\right)$. When the regularization mapping is the $\text{KCDF}$ function, the property is defined as:
\begin{equation}
\text{Sel}_{\text{KCDF}}\left(\alpha, \beta\right) := 1 - \int_{0}^1 \text{KCDF}\left(\omega; \alpha, \beta\right) d\omega\text{.}
\end{equation}
In essence, the selectivity property is a determiner of how selective the regularization mapping is; the higher the selectivity, the more picky the mapping about taking into account less important edges.


\section{Results}
\label{section:results}
\subsection{Data selection}
\label{subsection:data}
In this work, we chose the DREAM4 and DREAM5 networks as the base gold-standard networks from which the data sets were generated. The DREAM4 data sets \cite{marbach:2009dream4data} are available as part of the DREAM4 challenge \cite{stolovitzky:2007dream4challenge} and consist of five \textit{in silico} networks. Each network has 100 genes and is generated from specific gene modules chosen to mimic networks in two model species Escherichia coli and Saccharomyces cerevisiae. The DREAM5 networks are part of the DREAM5 standard challenge \cite{marbach:2012dream5challenge}. Two of the DREAM5 networks that we used in our experiments belong to the same model species, but are larger and are made based on curated interactions in E. coli and high-confidence set of interactions extracted from genome-wide transcription factor binding data (ChIP-chip) and evolutionarily conserved binding motifs in S. cerevisiae \cite{marbach:2012dream5challenge, fan:2018inferring}. In addition to the original DREAM5 networks, we also extracted some subnetworks these networks.

We used GeneNetWeaver (GNW) 3.0 \cite{schaffter:2011gnw} as the tool to generate the \textit{in silico} data sets from the gold-standard networks. For each network in the DREAM4 challenge, we produced 10 collections of training data. In each collection, there are three data sets: a multi-perturbation, a knockouts and a time-series data set were generated. All of the data sets were produced using the recommended settings in DREAM4. As in the DREAM4 challenge, internal noise based on stochastic differential equations and a mix of normal and log-normal noise as the measurement noise were added to each network. 

We used the same piece of software to extract sub-networks from the original networks in the DREAM5 challenge. In addition to the two original DREAM5 networks, this let us produce ten more networks of these two model species:
\begin{itemize}
    \item \textbf{E. coli sub-networks}: 5 sub-networks of the E. coli gold-standard network comprised of 40, 80, 160, 320, and 640 genes,
    \item \textbf{S. cerevisiae sub-networks}: 5 sub-networks of the S. cerevisiae gold-standard network comprised of 100, 200, 400, 800, and 1600 genes.
\end{itemize}
All of the above networks were generated using the GNW's network extraction functionality with its default settings, after removing the self-regulatory links. In the case of the S. cerevisiae sub-networks, the greater sub-networks are super-networks of the smaller subnetworks. For example, S. cerevisiae of size 1600 is a super-network of S. cerevisiae of size 800. We generated four data sets for each of the five networks in any group: A multi-factorial perturbation data set, a time-series data set, and two knockout data sets. All of these data sets were generated using GNW with the default settings.

Overall, there were 5 DREAM4 networks, 2 DREAM5 networks, and 10 DREAM5 sub-networks. We used each of these groups for distinct tests and evaluations. While the original networks will be used for the final evaluations, we will use the sub-networks as the validation sets, and extract the optimal parameters of GENEREF based on them.

\subsection{Performance evaluation}
\label{subsection:evaluation}
To evaluate the results from the algorithm, each final edge confidence matrix ($\hat{\mathbf{W}}_M$) is compared to the corresponding gold-standard network. Two evaluation metrics were considered: Area under the receiver operating characteristics (AUROC) and area under precision-recall curve (AUPR). The p-values of the metrics -- which are the probability that a random reconstruction algorithm produces an equal or greater value than the metric -- were calculated based on the methods introduced in \citen{stolovitzky:2009dream2}. Based on the p-values, the prediction performance metric for a single network can be defined as:
\begin{equation}
\textrm{score}\left(k\right) = -\frac{1}{2} \log_{10}{\left[p_\textrm{AUPR}{\left(k\right)} \times p_\textrm{AUROC}{\left(k\right)}\right]}
\end{equation}
where $p_\textrm{AUPR}{\left(k\right)}$ and $p_\textrm{AUROC}{\left(k\right)}$ are respectively the AUPR p-value of the $k$-th network and the AUROC p-value of the $k$-th network. The overall score of the algorithm over all $C$ networks is determined by averaging the scores obtained for each network:
\begin{equation}
\textrm{overall-score} = \frac{1}{C} \sum_{k=1}^{C}{\textrm{score}\left(k\right)}
\label{eq:score}
\end{equation}
Here, $C$ depends on the number of networks in the experiment. For example, if our experiment is performed on the five DREAM4 data sets, then $C = 5$.

\begin{figure}[t!]
\centering
\includegraphics[width=1.0\linewidth]{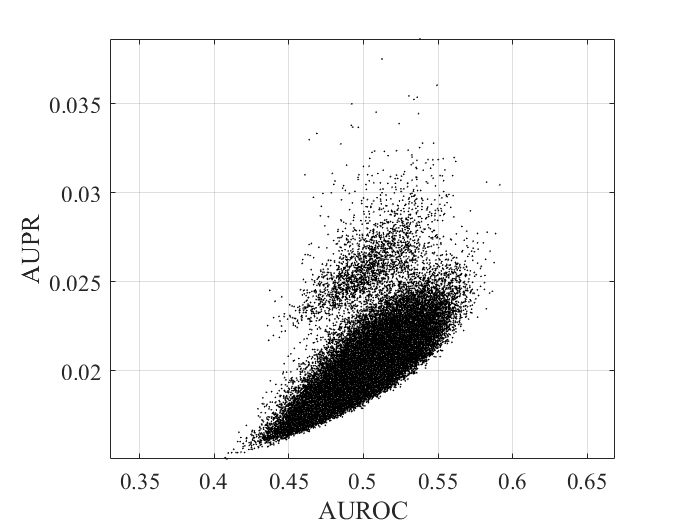}
\caption{The AUROC and AUPR of 100,000 evaluations of a random predictor used on a network with 100 genes and 176 edges (Network 1 from DREAM4).}
\label{fig:auroc-aupr}
\end{figure}

Our observations suggest that the AUROC and AUOR scores do not show uncorrelatedness to a confidential extent on sparse networks on the scale of DREAM4's. Refer to figure \ref{fig:auroc-aupr} as an example, where the random AUROC and AUPR values of Network 1 in DREAM4 are depicted and a lack of complete correltedness can be inferred. (We will further discuss the correlation between these two metrics in a network in Appendix \ref{sec:corr-coef}.) Moreover, it has been shown that the AUPR metric is more informative than the AUROC in classification problems with skewed class distributions as in biological regulatory networks and the DREAM4 and DREAM5 networks \cite{saito:2015skewedclasses}. Considering these two facts, we determined to introduce a second score metric too:
\begin{equation}
\textrm{score}_\textrm{AUPR}\left(k\right) = -\log_{10}{\left[p_\textrm{AUPR}{\left(k\right)}\right]}\textrm{,}
\end{equation}
and similar to equation \ref{eq:score}, the overall score for $\textrm{score}_\textrm{AUPR}$ will be calculated using the following equation:
\begin{equation}
{\textrm{overall-score}_\textrm{AUPR}} = \frac{1}{C} \sum_{k=1}^{C}{\textrm{score}_\textrm{AUPR}\left(k\right)}
\end{equation}



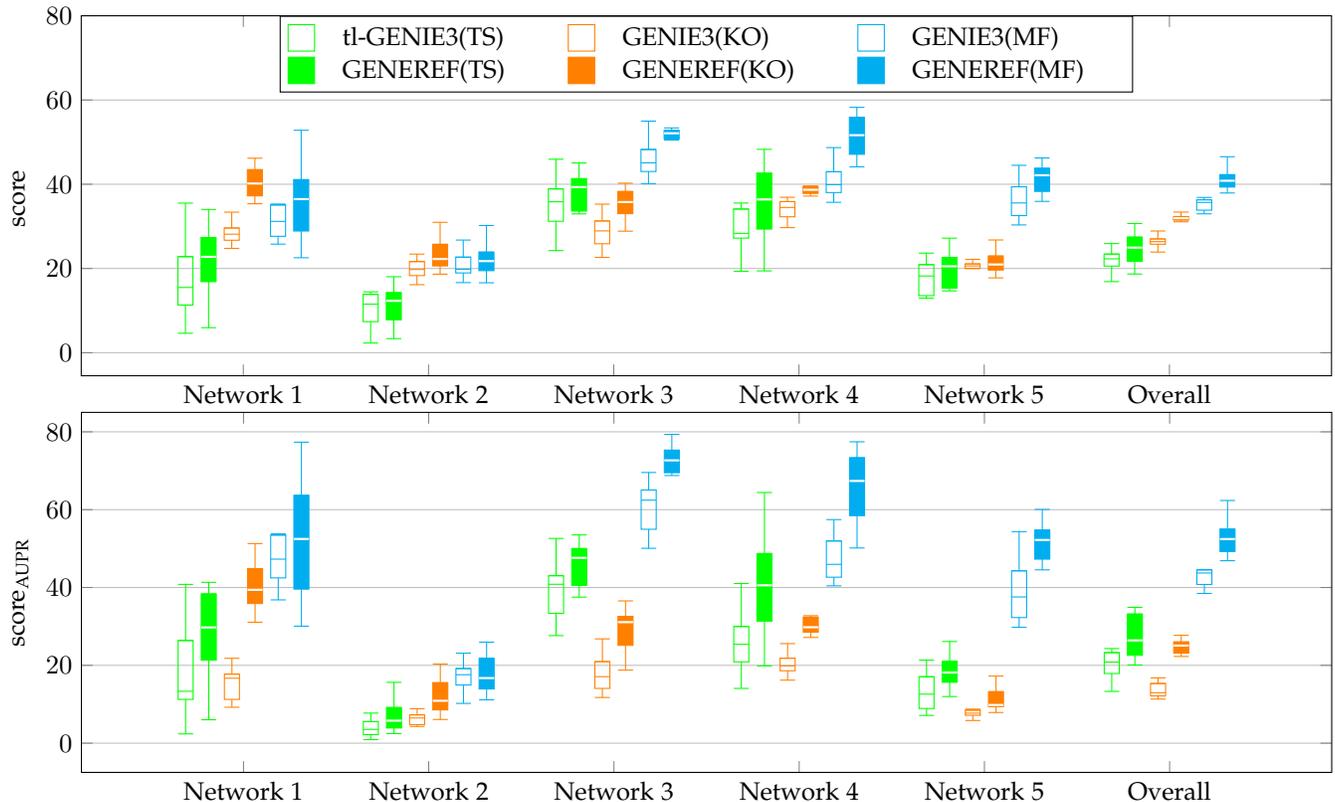
\begin{figure*}[t!]
\centering
\begin{tikzpicture}
\begin{axis} [box plot width=1mm, width=\linewidth, height=0.35\linewidth, xtick={-0.5, 7.5, 15.5, 23.5, 31.5, 39.5}, xticklabels={Network 1, Network 2, Network 3, Network 4, Network 5, Overall}, ylabel={$\text{score}_\text{\ }$}, ymajorgrids, legend cell align={left}, legend style={at={(0.5,1)},anchor=north ,column sep=0.3cm}, legend columns=3, ymax=80]
\boxplot [forget plot, green] {table [col sep=comma] {modulation/GENIE3_2,_score.csv}}
\boxplotfilled [forget plot, green] {table [col sep=comma] {modulation/GENEREF_2,_score.csv}}
\boxplot [forget plot, orange] {table [col sep=comma] {modulation/GENIE3_1,_score.csv}}
\boxplotfilled [forget plot, orange] {table [col sep=comma] {modulation/GENEREF_1,_score.csv}}
\boxplot [forget plot, cyan] {table [col sep=comma] {modulation/GENIE3_0,_score.csv}}
\boxplotfilled [forget plot, cyan] {table [col sep=comma] {modulation/GENEREF_0,_score.csv}}
\boxplotlegend[only marks,mark=square,green, mark options={mark size=5pt}]{{tl-GENIE3(TS)\ \ \ \ \ \ }}
\boxplotlegend[only marks,mark=square,orange, mark options={mark size=5pt}]{{GENIE3(KO)\ \ \ \ \ \ }}
\boxplotlegend[only marks,mark=square,cyan, mark options={mark size=5pt}]{{GENIE3(MF)\ \ \ \ \ \ }}
\boxplotlegend[only marks,mark=square*,green, mark options={mark size=5pt}]{{GENEREF(TS)\ \ \ \ \ \ }}
\boxplotlegend[only marks,mark=square*,orange, mark options={mark size=5pt}]{{GENEREF(KO)\ \ \ \ \ \ }}
\boxplotlegend[only marks,mark=square*,cyan, mark options={mark size=5pt}]{{GENEREF(MF)\ \ \ \ \ \ }}
\end{axis}
\end{tikzpicture}
\begin{tikzpicture}
\begin{axis} [box plot width=1mm, width=\linewidth, height=0.35\linewidth, xtick={-0.5, 7.5, 15.5, 23.5, 31.5, 39.5}, xticklabels={Network 1, Network 2, Network 3, Network 4, Network 5, Overall}, ylabel={$\text{score}_\text{AUPR}$}, ymajorgrids, legend cell align={left}, legend style={at={(0.5,1)},anchor=north ,column sep=0.3cm}, legend columns=-1, ymax=85]
\boxplot [forget plot, green] {table [col sep=comma] {modulation/GENIE3_2,_score_aupr.csv}}
\boxplotfilled [forget plot, green] {table [col sep=comma] {modulation/GENEREF_2,_score_aupr.csv}}
\boxplot [forget plot, orange] {table [col sep=comma] {modulation/GENIE3_1,_score_aupr.csv}}
\boxplotfilled [forget plot, orange] {table [col sep=comma] {modulation/GENEREF_1,_score_aupr.csv}}
\boxplot [forget plot, cyan] {table [col sep=comma] {modulation/GENIE3_0,_score_aupr.csv}}
\boxplotfilled [forget plot, cyan] {table [col sep=comma] {modulation/GENEREF_0,_score_aupr.csv}}
\end{axis}
\end{tikzpicture}
\caption{The effect of the modulation phase on the performance of the first iteration of GENEREF. 10 experiments for each model were done in order to obtain the robustness values.}
\label{fig:modulation-robustness}
\end{figure*}

\subsection{Evaluation of the competitive methods}
Among the competing algorithms are those that utilize only one type of data sets: SVR \cite{drucker:1997svr-algorithm} (which uses support vector regressors) and its ensemble variant E-SVR; TIGRES \cite{haury:2012tigress}, which applies the linear regression equation directly to the problem; GENIE3 \cite{irrthum:2010genie3} and its time-lagged variant tl-GENIE3, which are roughly equivalent to one iteration of GENEREF; BiXGBoost \cite{zheng:2018bixgboost}, which considers two local structure of GRNs into a bidirectional model and applies a gradient boosting with regressor trees \cite{chen:2016xgboost} to solve the model; and NIMEFI (GENIE3+E-SVR) \cite{ruyssinck:2014nimefi}, an ensemble method that combines predictions from GENIE3 and E-SVR. In order to have a fairer comparison, we also included dynGENIE3 \cite{geurts:2018dyngenie3}, which utilizes both types of data sets, and iRafNet \cite{petralia:2015irafnet}, which guides the training of a data set using an other data set. Both GENIE3 and dynGENIE3 are tree-based ensemble methods, with dynGENIE3 taking into account the ordinary differential equations acting in both the steady state and time series data sets. iRafNet, is another tree based algorithm that applies prior knowledge extracted from other data set during the training of random forests on the main data set. Unlike GENEREF, that alters the $\text{Improvement}$ function, iRafNet uses the prior weights to alter the distribution of feature selection probability during the construction of the trees.

The DREAM4 scores that we report for SVR, E-SVR, TIGRESS, and NIMEFI were taken from \citen{ruyssinck:2014nimefi}. 
We ran our own implementations for GENIE3, tl-GENIE3, and GENEREF. In order to obtain the results for dynGENIE3, the original implementation was run on our data sets. The scoring metrics of iRafNet were calculated using the AUROC and AUPR values reported in \citen{petralia:2015irafnet}. Also, we calculated the scoring metrics of BiXGBoost based on its AUROC and AUPR values reported in \citen{zheng:2018bixgboost}.

\subsection{Performance evaluation on the DREAM4 networks}
\label{subsection:dream4evaluation}

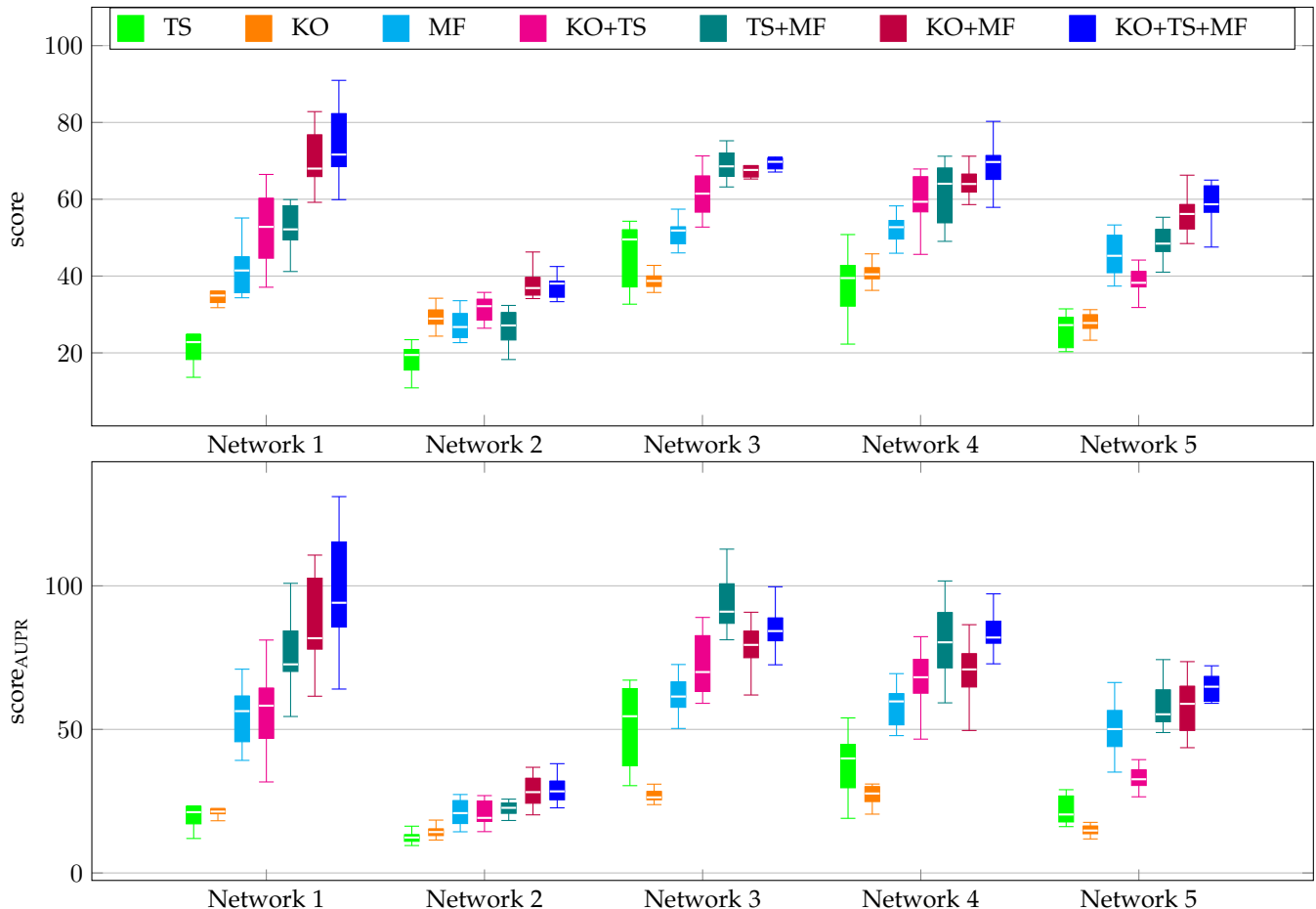
\begin{figure*}[t!]
\centering
\begin{tikzpicture}
\begin{axis} [box plot width=1mm, width=\linewidth, height=0.4\linewidth, xtick={0, 9, 18, 27, 36}, xticklabels={Network 1, Network 2, Network 3, Network 4, Network 5}, ylabel={$\text{score}_\text{\ }$}, ymajorgrids, legend cell align={left}, legend style={at={(0.5,1)},anchor=north ,column sep=0.2cm}, legend columns=-1, ymax=110]
\boxplotfilled [forget plot, green] {table [col sep=comma] {robustness/GENEREF_2,_score.csv}}
\boxplotlegend[only marks,mark=square*,green, mark options={mark size=5pt}]{TS\ \ \ \ \ \ }
\boxplotfilled [forget plot, orange] {table [col sep=comma] {robustness/GENEREF_1,_score.csv}}
\boxplotlegend[only marks,mark=square*,orange, mark options={mark size=5pt}]{{KO\ \ \ \ \ \ }}
\boxplotfilled [forget plot, cyan] {table [col sep=comma] {robustness/GENEREF_0,_score.csv}}
\boxplotlegend[only marks,mark=square*,cyan, mark options={mark size=5pt}]{MF\ \ \ \ \ \ }
\boxplotfilled [forget plot, magenta] {table [col sep=comma] {robustness/GENEREF_1,2_score.csv}}
\boxplotlegend[only marks,mark=square*,magenta, mark options={mark size=5pt}]{KO+TS\ \ \ \ \ \ }
\boxplotfilled [forget plot, teal] {table [col sep=comma] {robustness/GENEREF_2,0_score.csv}}
\boxplotlegend[only marks,mark=square*,teal, mark options={mark size=5pt}]{TS+MF\ \ \ \ \ \ }
\boxplotfilled [forget plot, purple] {table [col sep=comma] {robustness/GENEREF_1,0_score.csv}}
\boxplotlegend[only marks,mark=square*,purple, mark options={mark size=5pt}]{KO+MF\ \ \ \ \ \ }
\boxplotfilled [forget plot, blue] {table [col sep=comma] {robustness/GENEREF_1,2,0_score.csv}}
\boxplotlegend[only marks,mark=square*,blue, mark options={mark size=5pt}]{KO+TS+MF\ \ \ \ \ \ }
\end{axis}
\end{tikzpicture}
\begin{tikzpicture}
\begin{axis} [box plot width=1mm, width=\linewidth, height=0.4\linewidth, xtick={0, 9, 18, 27, 36}, xticklabels={Network 1, Network 2, Network 3, Network 4, Network 5}, ylabel={$\text{score}_\text{AUPR}$}, ymajorgrids, legend cell align={left}]
\boxplotfilled [forget plot, green] {table [col sep=comma] {robustness/GENEREF_2,_score_aupr.csv}}
\boxplotfilled [forget plot, orange] {table [col sep=comma] {robustness/GENEREF_1,_score_aupr.csv}}
\boxplotfilled [forget plot, cyan] {table [col sep=comma] {robustness/GENEREF_0,_score_aupr.csv}}
\boxplotfilled [forget plot, magenta] {table [col sep=comma] {robustness/GENEREF_1,2_score_aupr.csv}}
\boxplotfilled [forget plot, teal] {table [col sep=comma] {robustness/GENEREF_2,0_score_aupr.csv}}
\boxplotfilled [forget plot, purple] {table [col sep=comma] {robustness/GENEREF_1,0_score_aupr.csv}}
\boxplotfilled [forget plot, blue] {table [col sep=comma] {robustness/GENEREF_1,2,0_score_aupr.csv}}
\end{axis}
\end{tikzpicture}
\caption{The robustness evaluation of the $\text{score}$ and $\text{score}_\text{AUPR}$ metric using seven configurations of GENEREF: three 1-iteration, three 2-iteration, and one 3-iteration configurations.}
\label{fig:robustness_score_aupr}
\end{figure*}

As GENEREF can be launched with different arrangement of data sets, here we evaluate the influence of the number of the data sets and their types. We tested GENEREF with different combination of three data sets: a multi-factorial perturbation steady state data set, a multi-perturbation time-series data set, and a single knock-out steady state data set. For the sake of abbreviation, we call them respectively MF, TS, and KO. All data sets were generated using the GNW application, with the default settings. This means that all the data sets were generated with the same settings as the DREAM4 contest. We call an ordering of the data sets a configuration. Based on the number of data sets that GENEREF was trained on, the configurations that we used for our experiments can be divided into three categories:

\begin{itemize}
\item \textbf{GENEREF(MF)}, \textbf{GENEREF(TS)}, and \textbf{GENEREF(KO)}: On each of these configurations, only one data set was exploited. Note that the GENEREF algorithm in these configurations is equivalent to GENIE3 or tl-GENIE3 with the exception that GENEREF applies a modulation phase.
\item \textbf{GENEREF(KO+TS)}, \textbf{GENEREF(TS+MF)}, and \textbf{GENEREF(KO+MF)}: In this set of configurations, only two of the data sets were used. For example, GENEREF(TS+MF) trains on the MF data set, using the regularization matrix attained from the TS data set.

\item \textbf{GENEREF(KO+TS+MF)}: This configuration utilizes all the three data sets, in the indicated order, to infere the GRN.
\end{itemize}
Unless explicitly stated for an experiment, other orderings of the data sets were set aside.

There are two parameters in GENEREF that should be kept in consideration. They are the $\alpha$ and $\beta$ parameters that are the shape parameters of the mapping applied on the confidence matrices to obtain the feature importance matrices. Although the parameter can differ from one iteration in the algorithm to another, we chose to keep it constant throughout the algorithm for the sake of simplicity.

To eliminate the parameters and get an overall assessment for the performance that is comparable to present state-of-the-art methods, we performed a 5-fold cross validation on the five DREAM4 networks. To this end, for each of the five networks, we applied a grid search with $13$-by-$13$ logarithmically uniform intervals over the range of $\left[2^{-10}, 2^2\right]$ for $\alpha$ and $\left[2^{-4}, 2^8\right]$ for $\beta$ to retrieve an $\alpha$ and a $\beta$ value that maximized an assumed metric such as the AUROC. Then, for each of the five target networks, we averaged the optimal $\alpha$ and $\beta$ values of the other four networks and used these averaged values as the evaluation parameters to re-compute the metric on the target network. Finally, we reported the metric corresponding to the evaluation parameters on each of the target network as our algorithm's computed metric.

As our first experiment, we evaluated the influence of the modulation phase on GENEREF. The evaluation was done only using the first category of configurations, i.e., one-level GENEREF trained on the three data sets. To get a proper evaluation, we compared each GENEREF run with the corresponding configuration of the GENIE3 algorithm. Figure \ref{fig:modulation-robustness} shows the the comparison of the robustness of these two algorithms. In order to evaluate the robustness, the 10 collections of data sets for each of the 5 networks were used and the corresponding $\text{score}$ and $\text{score}_\text{AUPR}$ values were measured. The figure depicts an obvious improvement on every one of the 5 networks.

We also examined the robustness of our algorithm based on the number of iterations it was run. The same data sets were used. Figure \ref{fig:robustness_score_aupr} compares the robustness of all three categories of configurations. On each network, the first three box and whisker plots show the algorithm run on only one data set. The next three plots, depict the robustness of the two-dataset configurations of the algorithm. The last plot belongs to a run with three data sets. Although not universal, a general pattern of enhancement is observed when more data sets are added to the algorithm. Except for Network 2 with negligible improvement, all other data sets have a noticeable improvement when the third category of configurations is used in comparison to the second category. The improvement is in terms of both $\text{score}$ and $\text{score}_\text{AUPR}$.

\begin{figure*}[t!]
\begin{tikzpicture}
\begin{axis}[cycle list name=exotic, xtick={1,2,3}, width=0.45\linewidth, height=0.45\linewidth, ylabel={$\text{score}$}, xlabel={Number of iterations ($M$)}, ymajorgrids, legend cell align={left}]
\addplot table [x=x, y=network1, col sep=comma] {levels/plot_levels_score.csv};
\addplot table [x=x, y=network2, col sep=comma] {levels/plot_levels_score.csv};
\addplot table [x=x, y=network3, col sep=comma] {levels/plot_levels_score.csv};
\addplot table [x=x, y=network4, col sep=comma] {levels/plot_levels_score.csv};
\addplot table [x=x, y=network5, col sep=comma] {levels/plot_levels_score.csv};
\addplot[color=black,mark=*,line width=1.2pt, style=dashed] table [x=x, y=overall, col sep=comma] {levels/plot_levels_score.csv};
\end{axis}
\end{tikzpicture}
\begin{tikzpicture}
\begin{axis}[cycle list name=exotic, xtick={1,2,3}, width=0.45\linewidth, height=0.45\linewidth, ylabel={$\text{score}_\text{AUPR}$},xlabel={Number of iterations ($M$)}, ymajorgrids, legend cell align={left},legend pos=outer north east]
\addplot table [x=x, y=network1, col sep=comma] {levels/plot_levels_score_aupr.csv};
\addlegendentry{Network 1}
\addplot table [x=x, y=network2, col sep=comma] {levels/plot_levels_score_aupr.csv};
\addlegendentry{Network 2}
\addplot table [x=x, y=network3, col sep=comma] {levels/plot_levels_score_aupr.csv};
\addlegendentry{Network 3}
\addplot table [x=x, y=network4, col sep=comma] {levels/plot_levels_score_aupr.csv};
\addlegendentry{Network 4}
\addplot table [x=x, y=network5, col sep=comma] {levels/plot_levels_score_aupr.csv};
\addlegendentry{Network 5}
\addplot[color=black,mark=*,line width=1.2pt, style=dashed] table [x=x, y=overall, col sep=comma] {levels/plot_levels_score_aupr.csv};
\addlegendentry{Overall}
\end{axis}
\end{tikzpicture}
\caption{Average $\text{score}$ and $\text{score}_\text{AUPR}$ on the DREAM4 networks based on the number of iterations in GENEREF.  For $M \in \left\{1, 2, 3\right\}$ iterations, the average was taken on all $M$-permutations of the three data sets (the KO, TS, and MF data sets). All configurations with repetitive data sets have been dismissed.}
\label{fig:iteration_scores}
\end{figure*}
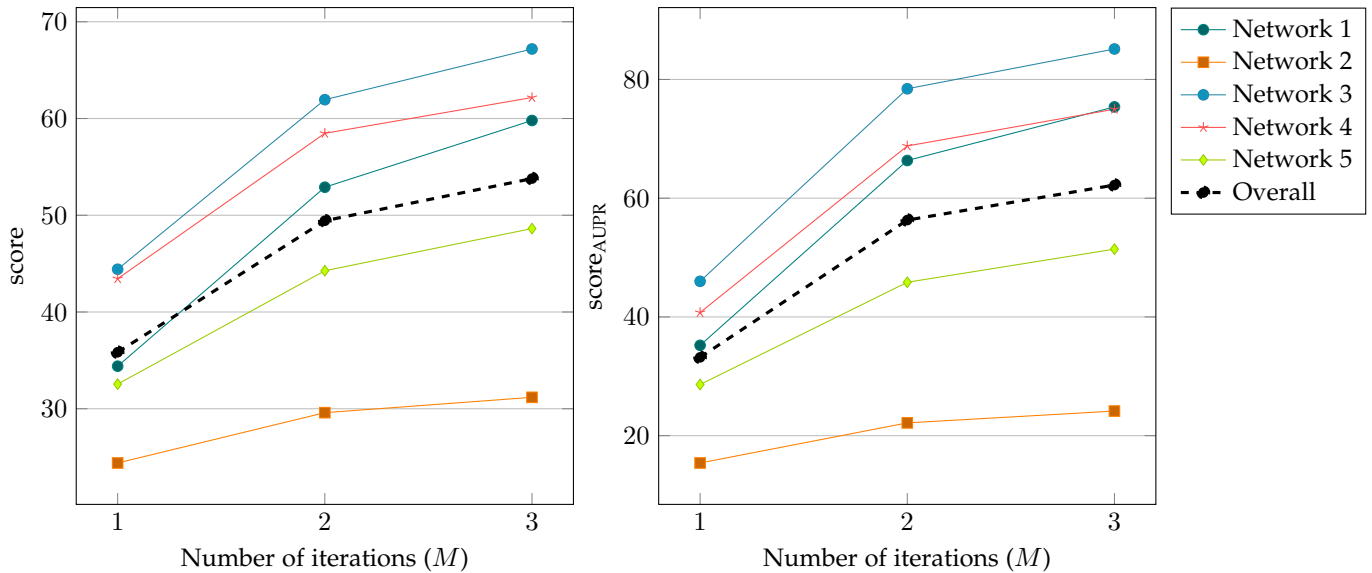

Along with the robustness evaluation, we directly tested the influence of the number of iterations on the performance of GENEREF. Figure \ref{fig:iteration_scores} compares the $\text{score}$ and $\text{score}_\text{AUPR}$ criteria of GENEREF versus the number of iterations. Here, instead of using the aforementioned categories, all $M$-permutations of the three data sets were tested for the $M$-iteration algorithm. This eliminates the effect of the order of data sets and gives an expectation of improvement based on $M$. An exceptionless improvement is observed as the number of iteration increses from $1$ to $3$.

To further evaluate our method, we also performed a Mann-Whitney test that compares four configurations of the algorithm with GENIE3 trained on the KO data set, GENIE3 trained on the MF data set, tl-GENIE3 trained on the TS data set, and dynGENIE3 trained on both the TS and MF data sets. These are the only algorithms that we managed to test directly. The results are shown in table \ref{tab:u-statistic}, which compares the $\text{overall-score}$ and $\text{overall-score}_\text{AUPR}$ metrics. As can be seen, the U-statistic marginally depend on the score metric. While GENEREF outperforms the other corresponding algorithms when using either score, the difference is more considerable when using the $\text{score}_\text{AUPR}$ metric. Also, based on the U-statistic, it can be concluded again that modulation phase has improved the overall performance of the algorithm.

For the last experiment, we compared the performance of various algorithms. We compared our algorithm to eight other methods: SVR, E-SVR, GENIE3, BiXGBoost, NIMEFI, iRafNet, and dynGENIE3. The results can be found on table \ref{tab:comparison}. To have a fair comparison, we only compared those configurations of GENEREF that utilized the same data set as the algorithms in comparison. The competing algorithms use one of the four groups of data sets. SVR, E-SVR, TIGRESS, NIMEFI and TIGRESS only use the MF data set. tl-GENIE3 works only based on the TS data set. iRafNet trains on the TS data set based on the information obtained from the KO data set. dynGENIE3 merges the MF and TS data sets and then performs the inference. 
None of the competing algorithms were run on all of the three data sets. Although the KO and MF data sets can hypothetically be combined to boost the performance, neither of the original papers have reported results based on a combination of these two data sets either. Therefore, the only algorithm using the three data sets and reported on the last group of data sets, is GENEREF.

\begin{table*}[!b]
\centering
\begin{tabularx}{\textwidth}{|>{\rule[-4pt]{0ex}{15pt}}X|X|X|}
\cline{2-3}

\multicolumn{1}{c|}{\hspace*{10cm}\quad} & \textbf{$\text{overall-score}$} & \textbf{$\text{overall-score}_\text{AUPR}$} \\
\cline{1-3} 
\hline \hline
tl\nobreakdash-GENIE3(TS)~vs.~GENEREF(TS) & 0.686 & 0.794 \\
\hline
GENIE3(KO)~vs.~GENEREF(KO) & 0.780 & 0.862 \\
\hline
GENIE3(MF)~vs.~GENEREF(MF) & 0.982 & 1.000 \\
\hline
dynGENIE3(TS\&MF)~vs.~GENEREF(TS+MF) & 0.648 & 0.768 \\
\hline
\end{tabularx}
\caption{\label{tab:u-statistic}The U statistic values derived from the Mann-Whitney test for comparison of GENEREF to the base algorithms. Number on a cell shows how probably the GENEREF algorithm performs better than the competing algorithm.}
\end{table*}

\begin{table*}[t!]
\centering
\begin{tabularx}{\textwidth}{|>{\bfseries}l|>{\bfseries}l|>{\bfseries}X||>{\rule[-1pt]{0ex}{12pt}}l|l|l|l|l|l|}
\hline
Metric & Data sets & Algorithm & Net 1\quad\quad & Net 2\quad\quad & Net 3\quad\quad & Net 4\quad\quad & Net 5\quad\quad & Overall \\
\cline{1-9} \hline \hline
\multirow{14}{*}{\rotatebox{90}{\textbf{$\text{score}$}\hspace{2cm}}} & \multirow{6}{*}{MF}
& SVR &  $2.46$ & $5.61$ & $6.76$ & $5.39$ & $3.08$ & $4.66$ \\
\cline{3-9}
\ & \ & E-SVR & $21.58$ & $\mathbf{45.76}$ & $41.16$ & $38.68$ & $37.30$ & $34.09$ \\
\cline{3-9}
\ & \ & TIGRESS & $26.70$ & $37.52$ & $41.03$ & $37.85$ & $36.65$ & $33.55$ \\
\cline{3-9}
\ & \ & NIMEFI & $27.45$ & $44.60$ & $48.15$ & $44.71$ & $44.60$ & $41.90$ \\
\cline{3-9}
\ & \ & GENIE3 & $27.17$ & $34.02$ & $44.30$ & $43.03$ & $38.21$ & $34.95$ \\
\cline{3-9}
\ & \ & GENEREF(MF) &  $\mathbf{32.56}$ & $36.14$ & $\mathbf{53.42}$ & $\mathbf{51.09}$ & $\mathbf{46.92}$ & $\mathbf{44.03}$ \\
\cline{2-9}\cline{2-9}
\hhline{|:~========}
\ & \multirow{4}{*}{TS} & BiXGBoost &  $20.54$ & $9.68$ & $24.30$ & $14.52$ & $16.54$ & $17.11$ \\
\cline{3-9}
\ & \ & tl-GENIE3 &  $22.50$ & $8.94$ & $30.32$ & $23.93$ & $21.27$ & 21.39 \\
\cline{3-9}
\ & \ & dynGENIE3(TS) &  $17.55$ & $9.16$ & $27.81$ & $\mathbf{39.78}$ & $15.67$ & $21.89$ \\
\cline{3-9}
\ & \ & GENEREF(TS) &  $\mathbf{30.08}$ & $\mathbf{10.11}$ & $\mathbf{39.96}$ & $34.42$ & $\mathbf{26.57}$ & $\mathbf{28.23}$ \\
\cline{2-9}\cline{2-9}
\hhline{|:~========}
\ & \multirow{2}{*}{KO, TS} & iRafNet &  $\mathbf{76.46}^*$ & $\mathbf{53.54}^*$ & $64.11$ & $57.47$ & $33.25$ & $\mathbf{56.97}$ \\
\cline{3-9}
\ & \ & GENEREF(TS+KO) &  $53.21$ & $38.67$ & $\mathbf{67.49}$ & $\mathbf{59.37}$ & $\mathbf{43.60}$ & $52.47$ \\
\cline{2-9}\cline{2-9}
\hhline{|:~========}
\ & \multirow{2}{*}{MF, TS} & dynGENIE3 & $39.75$ & $15.96$ & $54.49$ & $58.27$ & $38.36$ & $41.37$ \\
\cline{3-9}
\ & \ & GENEREF(TS+MF) &  $\mathbf{52.76}$ & $\mathbf{39.06}$ & $\mathbf{69.98}$ & $\mathbf{63.14}$ & $\mathbf{51.05}$ & $\mathbf{55.20}$  \\
\cline{2-9}\cline{2-9}
\hhline{|:~========}
\ & MF, TS, KO & GENEREF(TS+KO+MF) &  $\mathbf{73.61}$ & $\mathbf{38.21}$ & $\mathbf{71.55}^*$ & $\mathbf{73.43}^*$ & $\mathbf{59.26}^*$ & $\mathbf{63.21}^*$ \\
\cline{1-9}
\hline \hline
\multirow{14}{*}{\rotatebox{90}{\textbf{$\text{score}_\text{AUPR}$}\hspace{2cm}}} & \multirow{6}{*}{MF}
& SVR &  $2.46$ & $5.61$ & $6.76$ & $5.39$ & $3.08$ & $4.66$ \\
\cline{3-9}
\ & \ & E-SVR & $27.26$ & $52.92$ & $46.55$ & $41.13$ & $40.62$ & $41.70$ \\
\cline{3-9}
\ & \ & TIGRESS & $35.28$ & $53.96$ & $52.89$ & $47.62$ & $50.77$ & $48.10$ \\
\cline{3-9}
\ & \ & NIMEFI & $35.89$ & $\mathbf{56.85}$ & $58.70$ & $50.96$ & $52.80$ & $51.04$ \\
\cline{3-9}
\ & \ & GENIE3 & $37.34$ & $48.89$ & $61.00$ & $\mathbf{54.55}$ & $50.51$ & $50.46$ \\
\cline{3-9}
\ & \ & GENEREF(MF) &  $\mathbf{57.02}$ & $52.73$ & $\mathbf{72.01}$ & $43.88$ & $\mathbf{58.18}$ & $\mathbf{56.76}$ \\
\cline{2-9}\cline{2-9}
\hhline{|:~========}
\ & \multirow{4}{*}{TS} & BiXGBoost &  $27.49$ & $10.97$ & $38.13$ & $20.64$ & $22.63$ & $23.97$ \\
\cline{3-9}
\ & \ & tl-GENIE3 &  $13.27$ & $2.13$ & $26.99$ & $13.78$ & $13.00$ & $13.83$ \\
\cline{3-9}
\ & \ & dynGENIE3(TS) &  $11.92$ & $2.11$ & $24.77$ & $\mathbf{39.71}$ & $6.28$ & $16.95$  \\
\cline{3-9}
\ & \ & GENEREF(TS) &  $\mathbf{30.53}$ & $\mathbf{8.42}$ & $\mathbf{33.21}$ & $19.88$ & $\mathbf{28.03}$ & $\mathbf{24.14}$ \\
\cline{2-9}\cline{2-9}
\hhline{|:~========}
\ & \multirow{2}{*}{KO, TS} & iRafNet & $\mathbf{94.10}$ & $\mathbf{77.39}^*$ & $76.86$ & $74.29$ & $16.26$ & $\mathbf{67.78}$ \\
\cline{3-9}
\ & \ & GENEREF(TS+KO) &  $72.00$ & $44.76$ & $\mathbf{85.43}$ & $\mathbf{76.89}$ & $\mathbf{45.27}$ & $64.87$ \\
\cline{2-9}\cline{2-9}
\hhline{|:~========}
\ & \multirow{2}{*}{MF, TS} & dynGENIE3 & $38.25$ & $11.38$ & $72.77$ & $72.10$ & $36.76$ & $46.25$ \\
\cline{3-9}
\ & \ & GENEREF(TS+MF) &  $\mathbf{64.75}$ & $\mathbf{39.45}$ & $\mathbf{103.01}^*$ & $\mathbf{88.15}^*$ & $\mathbf{67.32}$ & $\mathbf{72.54}$  \\
\cline{2-9}\cline{2-9}
\hhline{|:~========}
\ & MF, TS, KO & GENEREF(TS+KO+MF) &  $\mathbf{99.21}^*$ & $\mathbf{41.11}$ & $\mathbf{94.18}$ & $\mathbf{85.12}$ & $\mathbf{76.51}^*$ & $\mathbf{79.23}^*$  \\
\hline
\end{tabularx}
\caption{\label{tab:comparison}Comparison of the performance of algorithms on the DREAM4 networks using various scoring metrics. The best results in each data set group are highlighted in boldface. The best total results are marked with an asterisk.}
\end{table*}

Two metrics were used for the comparisons: $\text{score}$ and $\text{score}_\text{AUPR}$. With any of the two metrics, GENEREF has the highest hit-count in every group of data sets. It also has the highest overall $\text{score}$ and $\text{score}_\text{AUPR}$ among all the algorithms.
When only one data set is used (MR or TS) GENEREF outperforms the competing algorithms on most of the networks. Even though it only applies a single improvement on GENEIE3 or tl-GENIE3, it turns out that this improvement has a drastic effect.
When the information from the MF and TS data sets is used GENEREF outperforms dynGENIE3 in all cases. iRafNet is the only algorithm that has higher values in more than one of the networks when the KO and TS have been used. It also has the higher overall performance in terms of both $\text{score}$ and $\text{score}_\text{AUPR}$. However, after the third data set (the MF data set) comes into play, the overall performance of GENEREF goes higher than that of iRafNet.

\subsection{Performance evaluation on the DREAM5 networks}
\label{subsection:dream5evaluation}
Our cross-validation tests on the DREAM4 networks suggest that ``independent networks of the same type and size \cite{stolovitzky:2007dream4challenge}'' lead to the same range of the optimal $\alpha$ and $\beta$ parameters. In this subsection, we will bring visual and systematic investigations of the optimal parameters of GENEREF. We start by evaluating the effect of three network characteristics of GRNs on the parameters: the type of the network (here, by comparing the variation of the parameters in E. coli versus S. cerevisiae -- a prokaryote specimen versus an eukaryote one), the size of the network, and also the number of iterations (number of data sets). After that, we proceed to evaluate GENEREF on the DREAM5 data sets and compare it to the competitive algorithms. 

\begin{figure*}[t!]
    \includegraphics[]{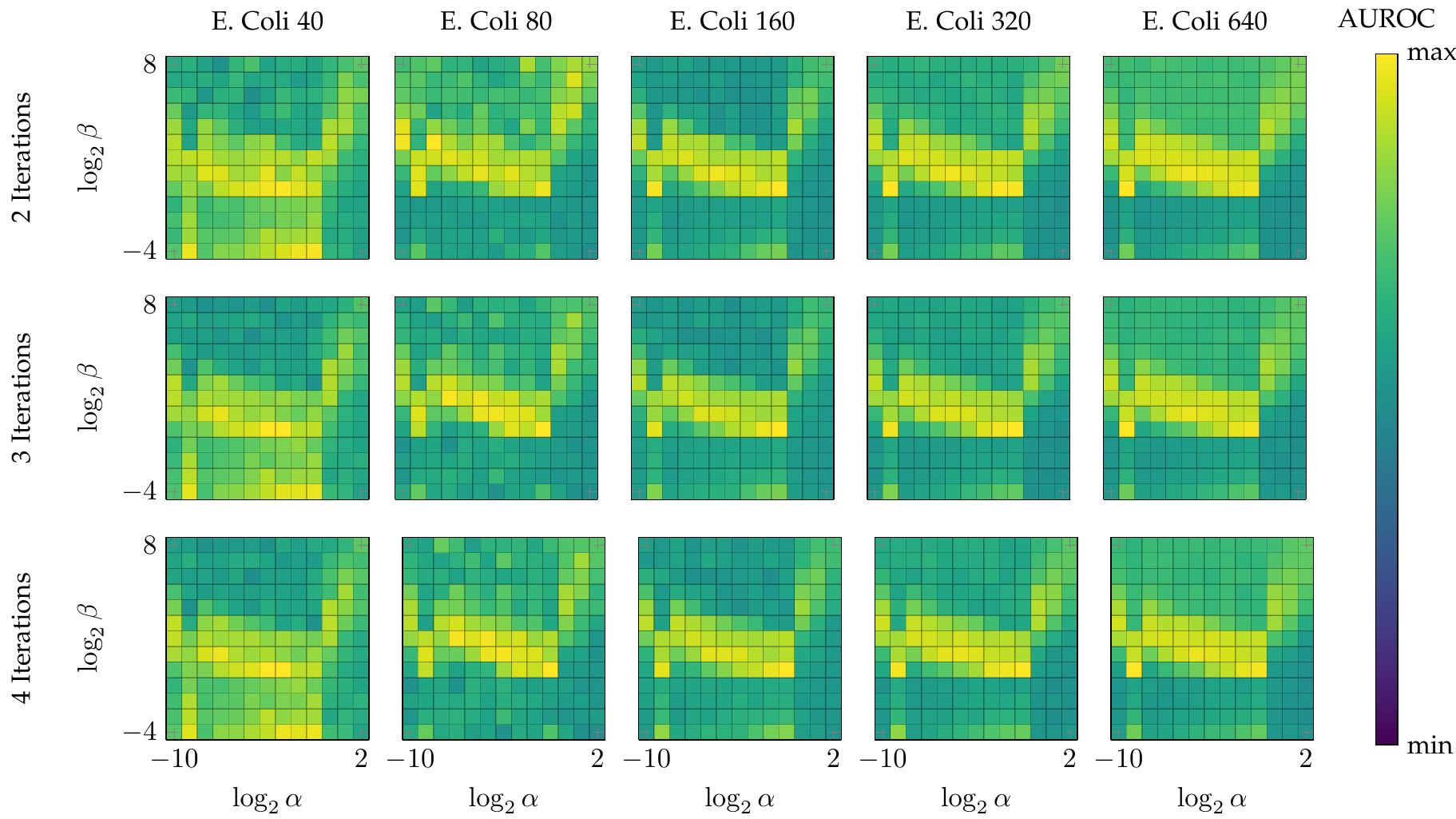}
    \\[0.3cm]
    \includegraphics[]{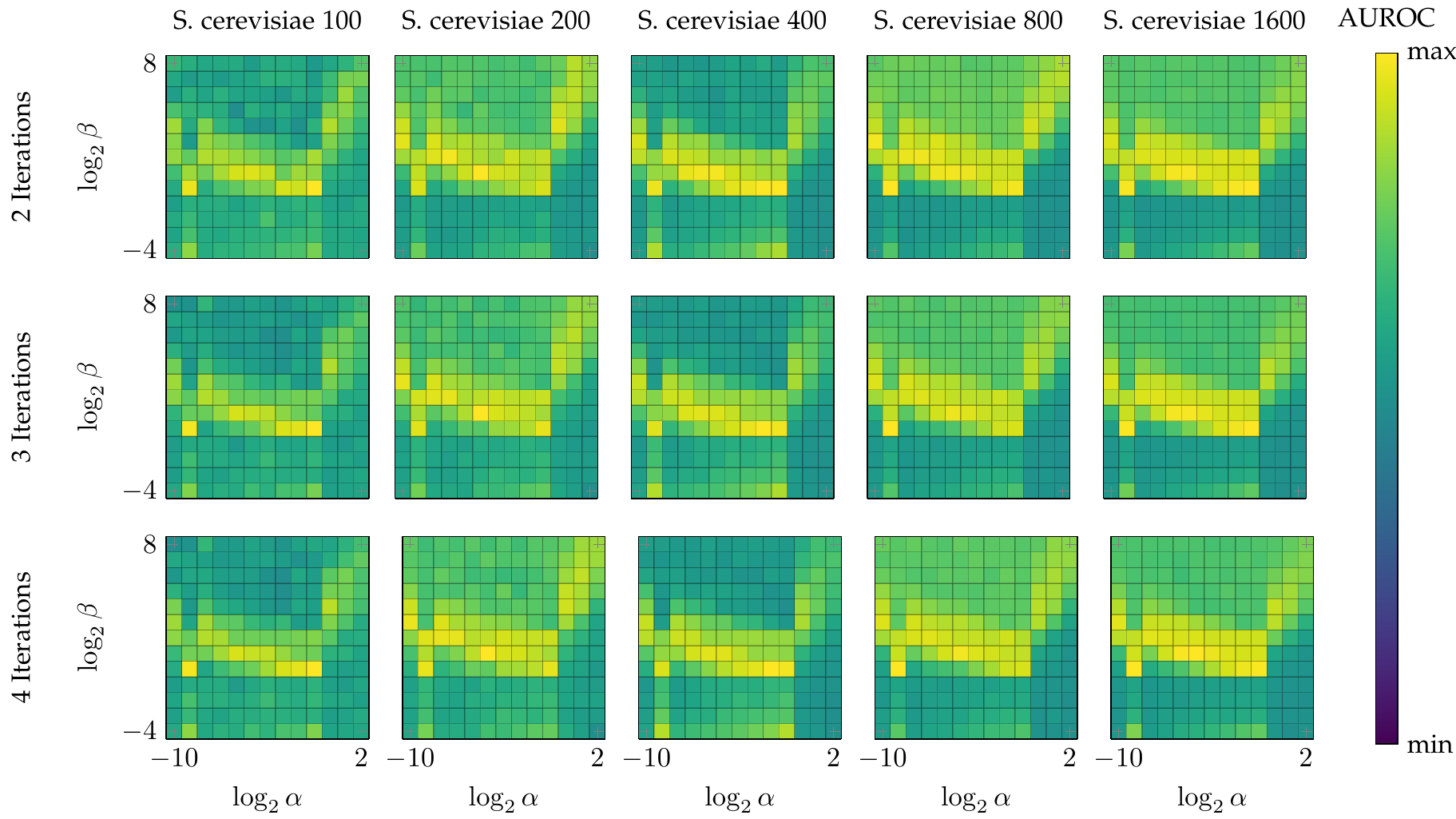}

\caption{The standardized AUROC metric as a function of the $\alpha$ and $\beta$ parameters. Five E. coli and five S. cerevisiae sub-networks were tested. Each color plot is scaled over the same range.}
\label{fig:generef_params_auroc}
\end{figure*}

\begin{figure*}[t!]
    \includegraphics[]{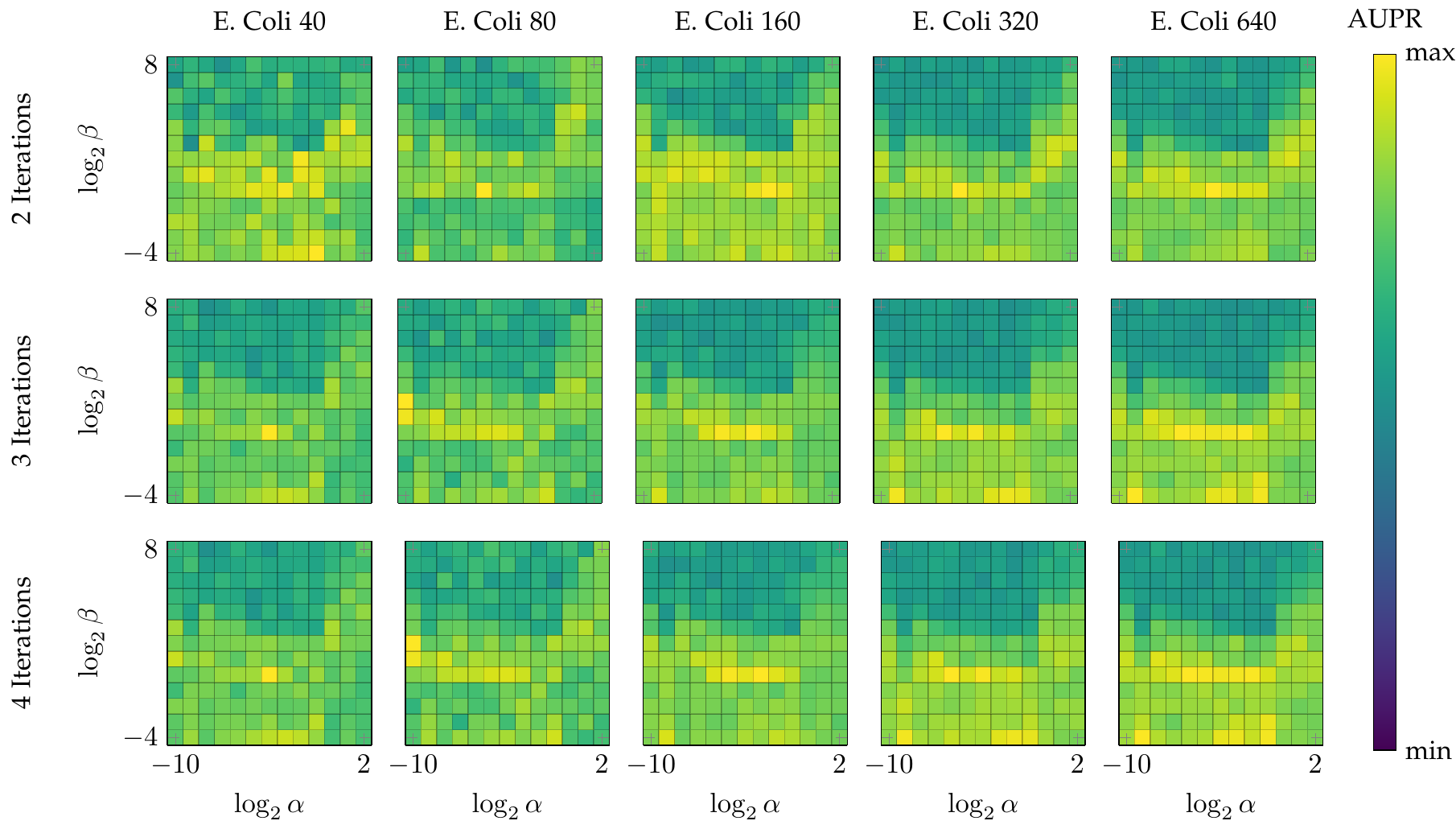}
    \\[0.3cm]
    \includegraphics[]{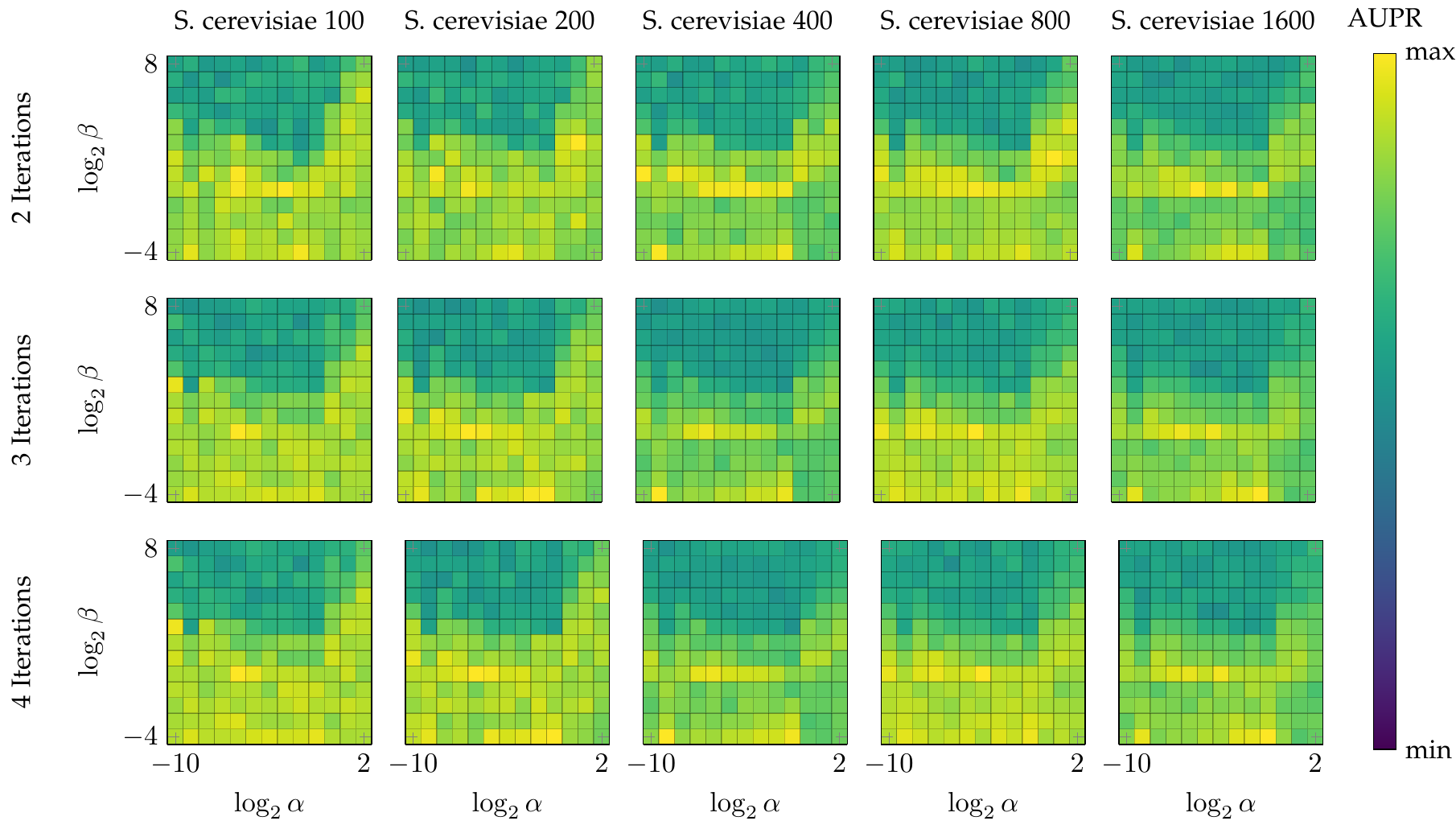}

\caption{The standardized AUPR metric as a function of the $\alpha$ and $\beta$ parameters. Five E. coli and five S. cerevisiae sub-networks were tested. Each color plot is scaled over the same range.}
\label{fig:generef_params_aupr}
\end{figure*}

We used the ten sub-networks that had been extracted from the DREAM5 original networks as the gold-standard networks of the first set of experiments. We presented the effect of the characteristics on the $\text{AUROC}$ metric in figure \ref{fig:generef_params_auroc}. This figure compares the GENEREF algorithm performance on each of the ten networks based on the number of iterations. Since the possible ordering of data sets for each number of iterations is arbitrary, we averaged all the permutations of the data sets that would constitute that number of iterations.
As we are only interested in the optimal range of parameters and not their corresponding metric values, we scaled all the plots to the same $\left[\text{min}, \text{max}\right]$ range to make the plots comparable. Each plot is the scaled AUROC value over $13\times13$ logaritmically uniform parameters in the range of $\left(\alpha, \beta\right) \in \left[\left(2^{-10}, 2^{-4}\right), \left(2^{2}, 2^{8}\right)\right]$. Figure \ref{fig:generef_params_aupr} shows the performance of the same networks when the $\text{AUPR}$ metric is used.

Based on an intuitive visual analysis of figures \ref{fig:generef_params_auroc} and \ref{fig:generef_params_aupr}, the effect of the size of the network on the range of the AUROC- and AUPR-maximizing $\alpha$ and $\beta$ values is not significant. The parameters approach their optimal values uniformly, with the uniformity becoming more robust as the size of network increases; 
there is no clear-cut difference between the plots of E. coli 320 and 640, or S. cerevisiae 800 and 1600. 
Neither does the addition of iterations influence the pattern drastically. The type of the network does not present a significant effect either; the difference between the patterns of either of the E. coli and S. cerevisiae is not considerable. Overall, the AUROC-maximizing parameter values ranged from $\left(\alpha, \beta\right) = \left(-5.33, 0.57\right)$ to $\left(\alpha, \beta\right) = \left(-4.00, 2.57\right)$, and the AUPR-maximizing parameter values ranged from $\left(\alpha, \beta\right) = \left(-4.65, -0.24\right)$ to $\left(\alpha, \beta\right) = \left(-3.75, 1.11\right)$ ($95\%$ confidence interval), with no visually evident effect on any of the mentioned network characteristics.

Besides the visual analysis, we analyzed the effects of the characteristics systematically. We will discuss these effects in terms of the \emph{optimal selectivity} of the algorithm, and define the optimal selectivity of GENEREF as the value of the selectivity function corresponding to the $\alpha$ and $\beta$ values that maximize an assumed metric -- typically $\text{score}$ or $\text{score}_\text{AUPR}$.

We expect the size of the network and number of iterations to have an effect on the optimal value of the parameters.
It is known that the average number of regulators per gene (denoted by $K$) remains roughly constant as the number of genes increases \cite{leclerc:2008sparse}. In other words, the edge density of the network ($E$) decreases by a factor of $g$:
\begin{equation}
    E = \frac{g \times K}{g \times (g-1)} \sim \frac{1}{g}
\end{equation}
This raises the hypothesis that the parameters should converge toward a higher optimal selectivity when networks become larger. 
Moreover, the algorithm is expected to have accumulated more information at higher iterations. If the accumulated information converge toward suggesting the same set of links, the optimal selectivity of the algorithm is expected to increase over iterations. On the other hand, if each new data set suggests novel links, the optimal selectivity may decrease. In general, we expect the optimal selectivity to be affected by the number of iterations too.
We tested these hypotheses by comparing changes in the size of the network to the changes of optimal selectivity in terms of AUROC and AUPR. To this end, we performed a p-value hypothesis testings on the group-wise correlation coefficients of three variables: the logarithm of the size of the network ($\log(g)$), the number of iterations ($M$), and the optimal selectivity of the network (optimal $\text{Sel}_\text{KCDF}\left(\alpha, \beta\right)$). We brought a rigorous explanation of these tests in appendix \ref{sec:optimal-params}. In sum, both of these tests affirmed that each of the network size and number of iterations have an effect on the optimal selectivity of the regularization mapping ($p < 10^{-7}$ and $p < 0.01$, respectively), although contrary to our expectation, the size of the network had a negative impact on the optimal selectivity when the AUPR metric was used.

\begin{table*}[t!]
\centering
\begin{tabularx}{\textwidth}{|>{\bfseries}l|>{\bfseries}l|>{\bfseries}X||>{\rule[-1pt]{0ex}{12pt}}l|l|l|}
\hline
Metric & Data sets & Algorithm & E. coli (Network 3)\quad\quad\quad\quad\quad\quad & S. cerevisiae (Network 4)\quad\quad & Overall \\
\cline{1-6} \hline \hline
\multirow{7}{*}{\rotatebox{90}{\textbf{$\text{score}$}}\hspace{1cm}} & \multirow{3}{*}{TS} & tl-GENIE3 &  $18.92$ & $16.52$ & $17.72$ \\
\cline{3-6}
\ & \ & dynGENIE3 & $\mathbf{23.43}$ & $\mathbf{28.91}$ & $\mathbf{26.17}$ \\
\cline{3-6}
\ & \ & GENEREF(TS) & $19.02$ & $16.83$ & $17.93$ \\
\cline{2-6}
\hhline{|:~=====}
\ & \multirow{2}{*}{MF/KO} & GENIE3 &  $41.12$ & $33.80$ & $37.46$ \\
\cline{3-6}
\ & \ & GENEREF(MF/KO) & $\mathbf{43.19}$ & $\mathbf{39.64}$ & $\mathbf{41.42}$ \\
\cline{2-6}
\hhline{|:~=====}
\ & \multirow{2}{*}{MF/KO, TS} & dynGENIE3 & $54.46$ & $\mathbf{72.12}^*$ & $\mathbf{63.29}^*$ \\
\cline{3-6}
\ & \ & GENEREF(MF/KO+TS) & $\mathbf{61.75}^*$ & $52.12$ & $56.94$ \\
\cline{1-6} \hline \hline

\multirow{7}{*}{\rotatebox{90}{\textbf{$\text{score}_\text{AUPR}$\hspace{1cm}}}} & \multirow{3}{*}{TS} & tl-GENIE3(TS) &  $9.82$ & $13.10$ & $11.45$ \\
\cline{3-6}
\ & \ & dynGENIE3 & $14.98$ & $\mathbf{21.29}$ & $\mathbf{18.14}$ \\
\cline{3-6}
\ & \ & GENEREF(TS) & $\mathbf{15.14}$ & $16.07$ & $15.61$ \\
\cline{2-6}
\hhline{|:~=====}
\ & \multirow{2}{*}{MF/KO} & GENIE3(MF/KO) & $12.82$ & $14.50$ & $13.66$ \\
\cline{3-6}
\ & \ & GENEREF(MF/KO) & $\mathbf{17.53}$ & $\mathbf{19.12}$ & $\mathbf{18.32}$ \\
\cline{2-6}
\hhline{|:~=====}
\ & \multirow{2}{*}{MF/KO, TS} & dynGENIE3 & $2.91$ & $18.00$ & $10.45$ \\
\cline{3-6}
\ & \ & GENEREF(MF/KO+TS) & $\mathbf{18.42}^*$ & $\mathbf{21.20}^*$ & $\mathbf{19.81}^*$ \\
\hline
\end{tabularx}
\caption{\label{tab:comparison-dream5}Comparison of the performance of algorithms on the DREAM5 networks. The best results in each metric group are highlighted in boldface and the best total results are marked with an asterisk. We omitted SVR and E-SVR from this table due to their low performance on the smaller DREAM4 data sets. Also, TIGRESS, NIMEFI, BiXGBoost, and iRafNet are excluded because they either did not report their results on DREAM5 or used other types of data like protein-protein interactions.}
\end{table*}

Table \ref{tab:comparison-dream5} reports the scores of various algorithms on the DREAM5 networks in comparison to competing algorithms. Here, we used the original learning data from the DREAM5 context as our data sets. Unlike the DREAM4 contest, DREAM5 has only one file consisting all of the expression levels of all of the experiments. To get the standard data sets, we dissociated each experiment thoroughly and formed two separate data sets; one by pooling all of the steady state profiles (MF/KO) and the other one by embedding all of the time-series profiles (TS). We performed a two-iteration GENREF on these two data sets.

Similar to DREAM4, we used cross-validation to obtain the optimal parameters for each of the DREAM5 networks. To get the expected optimal $\alpha$ and $\beta$ parameters for the E. coli network, we performed a linear regression on the S. cerevisiae sub-networks (including the original network itself) that maximized the considered metric in terms of the number of genes. We considered size of the network ($g$) as the input variable and the parameters as the output variables of the regression. We also set the number of iterations to $2$ (since were only two data sets) and only considered the optimal ordering of the data sets (MF/KO+TS for both $\text{score}$ and $\text{score}_\text{AUPR}$). We did the same procedure on the E. coli networks to obtain the expected optimal parameter values of S. cerevisiae.

Again GENEREF does not systemically surpass the other algorithms. Yet, when the algorithms are fed the whole learning data, GENEREF gives the best $\text{score}$ for the E. coli network and performs drastically better than the competitive algorithms when $\text{score}_\text{AUPR}$ is considered. When the MF/KO or TS data set is considered, GENEREF still improves GENIE3. However, dynGENIE3 is the best performer on the TS data set.

\section{Discussion}
In this paper, we introduced GENEREF, an iterative algorithm that can derive regulatory information between genes from various types of data sets. The evaluation was done on the synthetic and \textit{in vitro} data set networks. Although GENEREF does not systemically acquire the best performance when compared to other competitive algorithms, it has the highest overall scores when all the three DREAM4 data sets are used. Besides, GENEREF outperforms the competitive algorithms when performed on the E. coli data sets, if the $\text{score}$ metric is used and on the S. cerevisiae, if the $\text{score}_\text{AUPR}$ metric is used. Therefore, we conclude that GENEREF performs on par with other algorithms that utilize multiple data sets.

On the one hand, the diversity of existing data sets stems from the ongoing process of generating data from different kinds of experiments in laboratories around the world. Because of the widespread prevalence in production of data sets, it seems necessary to have advanced multi-dataset algorithms. Because our algorithm exploits the data sets in an incremental manner, it also has the advantage that new data sets can be fed to the algorithm without the need to reutilize the previously used data sets.

On the other hand, our observations reveal that, on our synthetic data sets, as the number of data sets increased, the improvement in the performance increase of adding the next data set plummeted. We hypothesize that this reduction can have been induced by at least one of these two scenarios: Firstly, if the data sets contain redundant information, the data sets that are on the close-to-final iterations will have less new information to reveal. The other problem that can arise when dealing with linear algorithms is that the patterns learned on previous iterations can start to fade out. In our case, the previously inferred regulatory links may vanish when new data sets are learned. This problem can be more prominent with the regulatory links that are detected only through surgical interventions (e.g. single knock-out/knock-down experiments). More experiments will be needed to reveal the exact cause of the plummeting improvement phenomenon.

It is recommended that data sets that can be merged are not considered as separate learning data and are fed to the algorithm as a single data set, even if multiple times. The recommendation applies to all multi-perturbation and knock-out data sets and time series data sets with the same configuration (e.g. those with identical time gaps and profiled in similar laboratory conditions). The reason is, as it is argued in \cite{deng:2013grrf}, in a tree node with a small number of instances, RRF is likely to select a feature not strongly relevant, and therefore performs poorly if the number of data records falls dramatically. In our tests on the DREAM5 networks, we obtained good results after coalescing all the steady-state and all of the time-series data sets, although the time gaps between the time-series data sets were not equal. However, further experiments to test the validity of this hypothesis might be required.

Although GENEREF principally outperforms both of the base algorithms, i.e. GENIE3 and GENIE3 (time-lagged), it cannot be concluded that it is the case for all GRNs. The results on the five DREAM4 networks are very data-dependent. The difference in the performance can be explained as a result of various aspects of the networks that control the regulatory relationship between the genes. We determined the number of genes as one of the regulating characteristics on the performance of the algorithm. Further investigations might be needed to determine other influential characteristics impacting the performance of the algorithm. Also, to investigate more deeply the performance of GENEREF, we plan to test it on a bigger real-world data set in the future.

It should be noted that there are ``combinatory methods'' that take into account the results of multiple algorithms and outperform the current generation of multi-dataset GRN reconstruction algorithms as with dynGENIE3. MCZ+dynGENIE3 \cite{geurts:2018dyngenie3} is one of these combinatory methods that makes use of the predictions of the two algorithms -- MCZ \cite{greenfield:2010mcz} and dynGENIE3 \cite{geurts:2018dyngenie3}. It multiplies the edge confidence matrices obtained from each of these algorithms. We did not discuss these combinatory methods throughout this paper and limited our comparisons to algorithms that generated the predictions independently. However, the usage of the predictions obtained from GENEREF in such combinatory methods can be a topic of research.

Similar to the AdaBoost algorithm  \cite{freund:1996adaboost}, on the conceptual level GENEREF can be thought of as machine learning meta algorithm that can exploit various regressors into a single model. Though, we used only random forests as our regressors, other regressor algorithms, including TIGRESS and dynGENIE3, can take this place. More generally, the data obtained from any other algorithm can be used to guide GENEREF's RRFs. Recently, methods based on deep learning and graph convolutional neural network methods have proven to be a great potential for GRN reconstruction and have taken a lot of attention \cite{zitnik:2018modeling, li:2019deep}. 
The combination and replacement of other regression algorithms or making use of information obtained from any of these advent approaches in the meta algorithm will be an other topic for exploration in our future work.

\section{Author contributions statement}
M.S. performed the experiments and analyzed the results.  M.S. and M.A. reviewed the manuscript. 

\section{Additional information}

Mr. Saremi declares no competing interests. Dr. AmirMazlaghani, who has supervised this thesis, has received compensation as an assistant professor at Amirkabir University of Technology for her supervision.

\ifCLASSOPTIONcompsoc
  \section*{Acknowledgments}
\else
  \section*{Acknowledgment}
\fi

The authors would like to thank Dr. Fatemeh Zare Mirakabad (Amirkabir University of Technology) who gave them continued support within the area of biology, during this research and the composition of this article.

\appendices
\section{Correlation Coefficient Between AUROC and AUPR}
\label{sec:corr-coef}

In order to have an evaluation of the dependency between the null distribution of both AUROC and AUPR, we calculated the correlation coefficient for the networks in both DREAM4 and DREAM5, as in table \ref{tab:dream-correlations}. In this table, $g$ represents the number of genes, and $K$ is the average number of regulators per gene. In none of these networks the $\text{AUROC}$ and $\text{AUPR}$ metrics demonstrate negligible linear dependency.

\begin{table}
\centering
\begin{tabularx}{\columnwidth}{|>{\rule[-1pt]{0ex}{12pt}}l|X|X|l|}
\cline{1-4}

\textbf{Network~Name}\hspace{0.5cm} & \textbf{$g$} & $K$ & \textbf{Correlation Coefficient} \\
\cline{1-4} 
\hline \hline
DREAM4 Network 1 & 100 & 1.76 & 0.8080 \\
\hline
DREAM4 Network 2 & 100 & 2.49 & 0.8242 \\
\hline
DREAM4 Network 3 & 100 & 1.95 & 0.8075 \\
\hline
DREAM4 Network 4 & 100 & 2.11 & 0.8171 \\
\hline
DREAM4 Network 5 & 100 & 1.93 & 0.8240 \\
\hline
DREAM5 E. coli & 1565 & 2.40 & 0.8755 \\
\hline
DREAM5 S. cerevisiae & 4441 & 2.90 & 0.9016 \\
\hline
\end{tabularx}
\caption{\label{tab:dream-correlations}The correlation between AUROC and AUPR of DREAM4 size 100 and DREAM 5 networks. $g$: number of genes; $K$: average number of transcriptional regulators per gene; Correlation Coefficient: the correlation between the AUROC and AUPR values of a random solver. Each correlation coefficient value is obtained from 100,000 randomly generated networks.}
\end{table}

\begin{table}
\centering
\begin{tabularx}{\columnwidth}{|>{\rule[-1pt]{0ex}{12pt}}l|X|}
\cline{1-2}

\textbf{Genes ($g$)}\hspace{1.5cm} & \textbf{AUROC-AUPR Correlation Coefficient} \\
\cline{1-2} 
\hline \hline
10 & $0.774 \pm 0.0025$ \\
\hline
20 & $0.740 \pm 0.0030$ \\
\hline
40 & $0.718 \pm 0.0038$ \\
\hline
80 & $0.723 \pm 0.0130$ \\
\hline
160 & $0.726 \pm 0.0046$ \\
\hline
320 & $0.732 \pm 0.0058$ \\
\hline
640 & $0.740 \pm 0.0083$ \\
\hline
1280 & $0.735 \pm 0.0124$ \\
\hline
2560 & $0.765 \pm 0.0150$ \\
\hline
5120 & $0.830 \pm 0.0211$ \\
\hline
10240 & $0.851 \pm 0.0417$ \\
\hline
\end{tabularx}
\caption{\label{tab:networks-correlation}The correlation between AUROC and AUPR of biological networks of different sizes. The
average number of transcriptional regulators per gene has been kept constant ($K = 1.875$).}
\end{table}

Furthermore, we demonstrated the effect of the network size on the correlation values in table \ref{tab:networks-correlation}. Note that the null AUROC and AUPR distributions only depend on the size number of links in the gold-standard network. We used a consistent edge density across all of these networks which we believe could represent that of real biological network. We averaged the values from table I in \cite{leclerc:2008sparse} of the following networks: two D. melanogaster networks, Sea urchin, S. cerevisiae, two S. cerevisiae networks, E. coli, and Arabidopsis thaliana.

It can be inferred from the tables that the number of actual edges ($gK$) has a positive impact on the correlation between the two metrics. However, the number of potential edges ($G (g-1)$) has compensated this effect. In none of the networks, the correlation coefficient value has fallen below $0.7$.

\section{The Relation Between Network Features and the Optimal Selectivity}
\label{sec:optimal-params}
In order to test the effect of the size of the network on the $\alpha$- and $\beta$-optimal selectivity of the regularization mapping we performed p-value hypothesis testings on the correlation coefficient between these two variables. We denoised the optimal parameters by taking the upper decile of each plot in figures \ref{fig:generef_params_auroc} and \ref{fig:generef_params_aupr} and averaging the corresponding $\alpha$ and $\beta$ values. Then computed the correlation coefficient values between these two variables grouped by the type of the network and the number of iterations. Table \ref{tab:size-selectivity-correlation} shows the results of the p-value tests on each of these groups.

\begin{table}
\centering
\begin{tabularx}{\columnwidth}{|>{\rule[-1pt]{0ex}{12pt}}l|l|X|p{1.5cm}|l|}
\hline

\textbf{Metric} & \textbf{Network} & \textbf{Number of\newline Iterations ($M$)} & \textbf{Correlation\newline Coefficient} & \textbf{p-value} \\
\cline{1-5} 
\hline \hline
\multirow{9}{*}{\rotatebox{90}{\textbf{$\text{AUROC}$\hspace{1cm}}}} & \multirow{4}{*}{E. coli} & 2 & 0.74 & 0.14 \\
\cline{3-5} 
\ & \ & 3 & 0.84 & 0.074 \\
\cline{3-5} 
\ & \ & 4 & 0.83 & 0.079 \\
\cline{3-5} 
\ & \ & -- & -- & 0.0053 \\
\cline{2-5} 
\ & \multirow{4}{*}{S. cerevisiae} & 2 & 0.40 & 0.49 \\
\cline{3-5} 
\ & \ & 3 & 0.84 & 0.068 \\
\cline{3-5} 
\ & \ & 4 & 0.82 & 0.087 \\
\cline{3-5} 
\ & \ & -- & -- & 0.014 \\
\cline{2-5} 
\ & -- & -- & -- & 0.00062 \\
\hline \hline
\multirow{9}{*}{\rotatebox{90}{\textbf{$\text{AUPR}$\hspace{1cm}}}} & \multirow{4}{*}{E. coli} & 2 & $-0.47$ & 0.42 \\
\cline{3-5} 
\ & \ & 3 & $-0.80$ & 0.10 \\
\cline{3-5} 
\ & \ & 4 & $-0.89$ & 0.037 \\
\cline{3-5} 
\ & \ & -- & -- & 0.0089 \\
\cline{2-5} 
\ & \multirow{4}{*}{S. cerevisiae} & 2 & $-0.96$ & 0.0079 \\
\cline{3-5} 
\ & \ & 3 & $-0.97$ & 0.0047 \\
\cline{3-5} 
\ & \ & 4 & $-0.95$ & 0.011 \\
\cline{3-5} 
\ & \ & -- & -- & 0.000061 \\
\cline{2-5} 
\ & -- & -- & -- & 0.000001 \\
\hline \hline
-- & -- & -- & -- & $5.4 \time 10^{-8}$ \\
\hline
\end{tabularx}
\caption{\label{tab:size-selectivity-correlation}The correlation between the logarithm of the number of genes ($\log(g)$) and the optimal selectivity of the regularization mapping (optimal $\text{Sel}_\text{KCDF}\left(\alpha, \beta\right)$). The last row of each group shows Fisher's combined p-value for all of the tests in that group.}
\end{table}

\begin{table}
\centering
\begin{tabularx}{\columnwidth}{|>{\rule[-1pt]{0ex}{12pt}}l|l|X|p{1.5cm}|l|}
\hline

\textbf{Metric} & \textbf{Network} & \textbf{Number of\newline Genes ($g$)} & \textbf{Correlation\newline Coefficient} & \textbf{p-value} \\
\cline{1-5} 
\hline \hline
\multirow{13}{*}{\rotatebox{90}{\textbf{$\text{AUROC}$\hspace{1.5cm}}}} & \multirow{6}{*}{E. coli} & 40 & $-0.86$ & $0.33$ \\
\cline{3-5} 
\ & \ & 80 & $-0.88$ & $0.30$ \\
\cline{3-5} 
\ & \ & 160 & $-0.85$ & $0.34$ \\
\cline{3-5} 
\ & \ & 320 & $-0.88$ & $0.31$ \\
\cline{3-5} 
\ & \ & 640 & $-0.83$ & $0.37$ \\
\cline{3-5} 
\ & \ & -- & -- & $0.054$ \\
\cline{2-5} 
\ & \multirow{6}{*}{S. cerevisiae} & 100 & $-0.86$ & $0.33$ \\
\cline{3-5} 
\ & \ & 200 & $-0.90$ & $0.28$ \\
\cline{3-5} 
\ & \ & 400 & $-0.94$ & $0.21$ \\
\cline{3-5} 
\ & \ & 800 & $-0.85$ & $0.34$ \\
\cline{3-5} 
\ & \ & 1600 & $-0.92$ & $0.24$ \\
\cline{3-5} 
\ & \ & -- & -- & $0.030$ \\
\cline{2-5} 
\ & -- & -- & -- & $0.0091$ \\
\hline \hline
\multirow{13}{*}{\rotatebox{90}{\textbf{$\text{AUPR}$\hspace{2cm}}}} & \multirow{6}{*}{E. coli} & 40 & $-0.77$ & $0.43$ \\
\cline{3-5} 
\ & \ & 80 & $-0.88$ & $0.31$ \\
\cline{3-5} 
\ & \ & 160 & $-0.55$ & $0.62$ \\
\cline{3-5} 
\ & \ & 320 & $-0.51$ & $0.65$ \\
\cline{3-5} 
\ & \ & 640 & $-0.87$ & $0.32$ \\
\cline{3-5} 
\ & \ & -- & -- & $0.12$ \\
\cline{2-5} 
\ & \multirow{6}{*}{S. cerevisiae} & 100 & $-0.57$ & $0.61$ \\
\cline{3-5} 
\ & \ & 200 & $-0.17$ & $0.89$ \\
\cline{3-5} 
\ & \ & 400 & $-0.71$ & $0.48$ \\
\cline{3-5} 
\ & \ & 800 & $-0.34$ & $0.77$ \\
\cline{3-5} 
\ & \ & 1600 & $-0.62$ & $0.57$ \\
\cline{3-5} 
\ & \ & -- & -- & $-0.33$ \\
\cline{2-5} 
\ & -- & -- & -- & $-0.15$ \\
\hline \hline
-- & -- & -- & -- & $0.0089$ \\
\hline
\end{tabularx}
\caption{\label{tab:iter-selectivity-correlation}The correlation between the number of iterations ($M$) and the optimal selectivity of the regularization mapping (optimal $\text{Sel}_\text{KCDF}\left(\alpha, \beta\right)$). The last row of each group shows Fisher's combined p-value for all of the tests performed in that group.}
\end{table}

We performed the same procedure to test the effect of the number of iterations and the optimal selectivity of the regularization mapping. Results are shown in table \ref{tab:iter-selectivity-correlation}.

\ifCLASSOPTIONcaptionsoff
  \newpage
\fi



%

%


\begin{IEEEbiography}[{\includegraphics[width=1in,height=1.25in,clip,keepaspectratio]{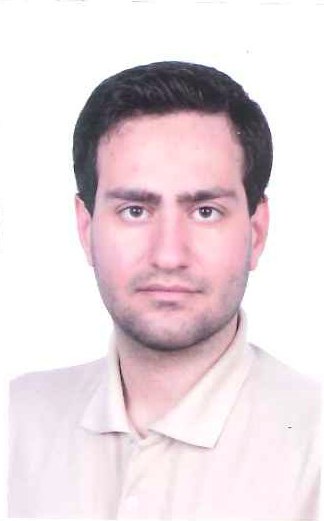}}]{Mehrzad Saremi}
is an M.Sc. graduate from Amirkabir University of Technology in Artificial Intelligence. He received his M.Sc. in 2019. His research interests revolve around Bioinformatics and Causal Graph Models.
\end{IEEEbiography}


\begin{IEEEbiography}[{\includegraphics[width=1in,height=1.25in,clip,keepaspectratio]{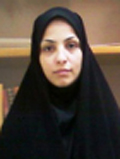}}]{Maryam Amirmazlaghani}
received the Ph.D. degree in electrical engineering from the Amirkabir University of Technology, Tehran, in 2009. She is currently a Faculty Member with the Department of Computer Engineering and Information Technology, Amirkabir University of Technology.

Her research interests include machine learning, optimization, statistical modeling and their applications.
\end{IEEEbiography}




\end{document}